# Is there something missing from the Antikythera Mechanism? Was it a mechanical Planetarium – positioner? or a Luni-solar Time calculator device?
## (Reconstructing the lost parts of b1 gear and its Cover Disc)

**Aristeidis Voulgaris[1], Christophoros Mouratidis[2], Andreas Vossinakis[3], Manos Roumeliotis[4]**

[1] City of Thessaloniki, Directorate Culture and Tourism, Thessaloniki, GR-54625, Greece
e-mail: arisvoulgaris@gmail.com (corresponding author)
[2] Merchant Marine Academy of Syros, GR-84100, Greece
[3] Thessaloniki Astronomy Club, Thessaloniki, GR-54646, Greece
[4] University of Macedonia, Department of Applied Informatics, Greece

**Abstract**

*We present the observations and the results of our experience from a large number of "flight hours" in constructing, assembling, handling, testing, studying and interacting with our Antikythera Mechanism functional reconstruction models. We constructed our models according to the X-Ray tomographies and without any use of modern stabilization parts (screws, nuts etc.). Simple typical bronze material (not with special alloys) was used in the construction, while (most of) the gears were cut from simple bronze discs with thickness ≈2mm and with triangular shaped teeth (not in modern involute shape). The bronze (not iron or steel) shafts have dimensions conforming to the original parts of the ancient artifact. The assembly of the parts was done following careful study of the "Personal Constructional Characteristics" and the Design Style of the ancient Craftsman, which are visible or even hidden on the Antikythera Mechanism Fragments. During the extensive use of our models, it was concluded that two important and mandatory indicators are missing from all current reconstructions of the Antikythera Mechanism. The absence of these two indicator dials makes it difficult to properly operate the Mechanism and their existence is necessary, in order for the Antikythera Mechanism to be considered as a complete and self-contained operational time-measuring device. The two procedures related to the (poorly) preserved remains were located on gear b1 and the lost Cover Disc of the b1 gear. The reconstruction of those missing parts was achieved by paying special attention to the Personal Constructional Characteristics of the ancient Craftsman. The extensive analysis of the Antikythera Mechanism's calibrated scales leads to the understanding of the Mechanism as a luni-solar time measuring device, as opposed to the notion that it was a mechanical planetarium presenting the hypothesized planetary motions and positions.*

## 1. Introduction

After the extraction of the Antikythera Mechanism from the sea bottom of the Antikythera gulf during the underwater excavation of 1900, and its discovery as a *strange mechanism with gears and inscriptions* in 1901/1902, many opinions were expressed about the operation of this remarkable device. The general assumption in the 1920s was that it was an astrolabe for ship navigation or some related operation (Theofanidis 1929). Through time and after the discovery and the recognition of new mechanical and inscriptional findings, the idea of an astrolabe was abandoned in favor of a mechanical calendar/astronomical calculator. The partially/poorly preserved engraved inscriptions with the names of the five known planets during antiquity, brought forward the hypothesis that this geared device was a mechanical planetarium, which should have had spheres and pointers on its front central dial, to represent the planets' motion and their position in the Ecliptic (Price 1974; Wright 2002; Freeth and Jones 2012; Freeth et al., 2021).

In 2004, following the initiative of M.Edmunds, J.H.Seiradakis and X.Moussas, new data were collected during the *Antikythera Mechanism Research Project*, leading to better knowledge about this geared machine (Freeth et al., 2006 and 2008).

One of the most characteristic procedures of the Antikythera Mechanism is the eclipse prediction via the eclipse scale (Saros spiral) and its pointer: Engraved on some of the 223 cells of the Saros spiral, there are solar/lunar eclipse events consisting the eclipse event sequence (Freeth et al., 2008; Anastasiou et al., 2016; Freeth 2019; Iversen and Jones 2019; Voulgaris et al., 2023b). The Antikythera Mechanism was designed and constructed with the ability to perform timed calculations and predict eclipse events and the time of their occurrence without needing any external information (see Voulgaris et al., 2023b, p. 22-24).

## 2. The measuring units of the Antikythera Mechanism

The Antikythera Mechanism was a computing device capable of carrying out complex calculations. The calculated results were presented via several pointers, which rotated around calibrated dials (i.e. scales with subdivisions). The motion of the Mechanism's pointers was achieved by the engagement of a number of gears, each having a specific teeth number related to characteristic astronomical numbers e.g. 223 teeth in gear $e_3$ (related to the 223 synodic months = one Saros), the engaged gears $d_2$, $d_1$ and $c_1$ = 127*(48/24) = 254 (related to the Metonic cycle of 254 sidereal months), etc.

Today, four measuring dials are preserved, in a circular distribution (Zodiac month ring, Egyptian calendar ring, Athletic Games circle, Exeligmos circle) which are divided in equal dimension subdivisions and two dials in spiral distribution, the Metonic and the Saros spirals, which are divided in constant angular frequency.

*What type of measurements did the Antikythera Mechanism perform?*
The answer can be given by checking the measuring units and subdivisions engraved on their corresponding dials. E.g. by observing the units on the measuring scale of a multimeter, someone can figure out that this measuring instrument measures Volts, Amperes and Ohms. These units are directly correlated to the nature of the electric current.
*The concept of relevance* governs a measuring machine: i.e. the procedures of the measurements are made on inter-correlated phenomena: it is difficult to find a real multimeter for electric current measurements and at the same time this device to offer the ability to measure distance in kilometers or the intensity of an earthquake in Richter's scale.
According to **Table 1**, all currently preserved scales of the Antikythera Mechanism measure time in units of days, synodic months and years. Each measuring scale is subdivided in equal time duration.
Based on the characteristics of these units, the Antikythera Mechanism is related to the procedures of a *Mechanical Time Measuring Calculator*.

**Table 1:** Preserved dials of the Antikythera Mechanism and their corresponding units. The type of measurement for each scale is the same: Measurement of Time.

| Measuring dial | Antikythera Mechanism Scale | Scale Subdivisions/Units | Type of measurement |
|---|---|---|---|
| Egyptian calendar ring | Egyptian year of full 365 Days | 1 Day | Time |
| Zodiac month ring | 1 Tropical year of 365.25 Days (365 equal subdivisions + 0.25 or 365.0 subdivisions with a correction procedure)[1] | 1 Day | Time (*instead of space in arc degrees*) |
| Metonic spiral | 19 Years/235 Synodic months | 1 Synodic month/cell | Time |
| Saros spiral | 18.03 Years/223 Synodic months | 1 Synodic month/cell | Time |
| Exeligmos dial | 54. 09 = 3 × Saros (= 3 × 223 Synodic months) | 18.03 Years/sector | Time |
| Athletic Games dial | 4 Years (= 49 or 50 Synodic months) | 1 Year/quadrant | Time |
| Parapegma plates | 1 Year = 4 × 1 (corner) columns 1 column = 1 Season = 3 month Each Parapegma event corresponds to 1 specific Date = 1 Day per index number | 1 Day | Time |

## 3. Present day missing parts of the Antikythera Mechanism

Today, 7 large and 75 smaller fragments of the Mechanism are preserved as shown in **Figure 1**. These fragments comprise a percentage of the Mechanism. By applying the Symmetry on the Design, the dimension and the shape of the Back and Front plates can be calculated and the partially preserved parts' can be reconstructed (Allen et al., 2016; Voulgaris et al., 2019b and 2021).

The reconstruction of the Mechanism's Back plate is based on the preserved fragments A, B, F and E and the Front plate is based on the remaining parts of fragments A, C, and a large number of small fragments. The mid-vertical line is the vertical axis of Symmetry, which crosses the preserved axis $b_{in}$/center of gear b1 (geometrical center of the Front plate).

Most of the Front Central Face parts of the Mechanism are missing. The only preserved part is the Lunar Disc (or Lunar Cylinder) located on Fragment C **Figure 2**. Close to the perimeter of the Lunar Cylinder a cavity is preserved, which was created by the lost Lunar Phases sphere (Wright 2006; Carman and DiCocco 2016). The lost Lunar phases sphere was probably made of an organic material, such as wood or ivory. The axis-z and the gear-z with 20 teeth, are the two parts (preserved by half) of the Lunar Phases sphere gearing (Wright 2006; Freeth et al., 2006; Carman and DiCocco 2016; Voulgaris et al., 2018b).

---

[1] *It was also suggested to divide the Zodiac ring in 360 un-equal subdivisions, i.e. arc degrees (units of space) (Evans, Carman and Thorndike 2010) in order to represent the solar anomaly, but the Parapegma events occurred in days, not in degrees (Anastasiou et al., 2013; Bitsakis and Jones 2016a; Lehoux 2012; Manitius 1880). Therefore, the Parapegma index letters which are engraved on some of the Zodiac ring subdivisions should correspond to days (equal-subdivisions), instead of arc-degrees (un-equal subdivisions). The ideal condition is the division of the Zodiac ring in 365 subdivisions + 0.25 subdivision, as similar division must have applied for the lost, but very probable Lunar day scale of 29 subdivisions + 0.5 subdivision (=29.5 days) (see **Figure 17** and **18**). By dividing the Zodiac ring in 365.25 (or rounded in 365.0) equal subdivisions and in different number of subdivisions per each month, the solar anomaly is represented on the Antikythera Mechanism (Voulgaris et al., 2018a).*

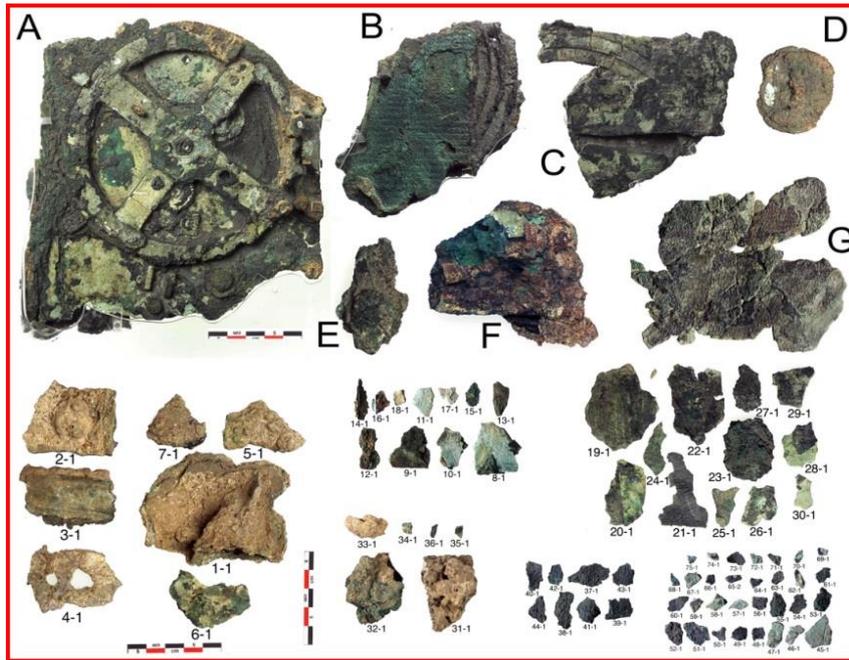

***Figure 1.*** *The Fragments of Antikythera Mechanism shown to scale). Above, the larger Fragments A-G. Below, the smaller fragments 1-75. The fragments are exhibited in the National Archaeological Museum of Athens, Greece (inventory number X.15087). Photos by K.Xenikakis, Copyright ©Hellenic Ministry of Culture & Sports/ Archaeological Receipts Fund.*

As inferred by the well preserved shape of the Lunar Cylinder surface, the specific position of the gear-z is the original position of the gear (i.e. the gear teeth aiming towards the perimeter of the Lunar cylinder, Carman and DiCocco 2016) and it cannot be adapted in inverted position (in Wright 2006; Freeth and Jones 2012; Freeth et al., 2021 the gear-z is presented in inverted position, i.e. the gear teeth aiming toward to the center of the Lunar Disk). Taking into account the original position of the gear z, this means that the Lunar phases sphere gearing needs three additional hypothetical gears and one shaft in order to operate (Voulgaris et al., 2018b; see also gearing scheme of Voulgaris et al., 2019b).

If gear-z could be placed in inverted position, then only one additional gear would be necessary for Lunar sphere rotation and not two additional gears. But the specific position of gear z seems to be the original since it cannot be adapted in the inverted position.

The question arises "*Why didn't the Craftsman of the AM invert the gear position to avoid the construction and assembly of two additional gears and their shaft?*"

The imprint in horseshoe shape at the internal area of the Lunar cylinder implies that there existed a spacer in horseshoe design, attached between the Lunar cylinders' cap and its, missing today, base.

As the base of the Lunar Cylinder is not preserved, a new question arises: *Could the original design of the Lunar Cylinder and its gearing be different than present day hypotheses?*

The lack of answers leaves these two questions open today...

There are also some poorly preserved mechanical formations on gear b1 (presented further below), remains from a mechanical structure just above gear b1, with unknown operation and design, and totally missing today (also discussed below).

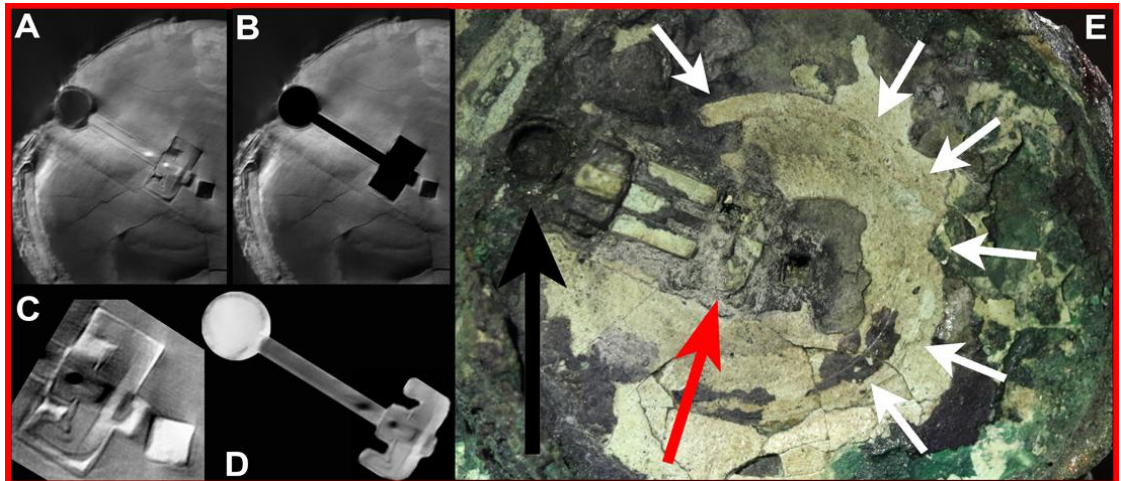

*Figure 2. A)* X-ray tomography of the Lunar Cylinder gear z, its axis and the Lunar Phases sphere cavity, located on Fragment C. *B)* The outline of the area in which the (lost) Lunar phases sphere, the gear z and its axis are attached. *C)* A close-up of the gear z. Note that the outline of this area precisely follows the outline of the gear. Attaching gear z in inversed position (and by keeping aligned the pin hole in the gear and in the axis), cannot be realistic (see also Carman and DiCocco 2016). The CT-scan slices were aligned to the Lunar cylinder surface and were further processed by the authors using the Real3D VolViCon software. *D)* Gear z, its axis and the Lunar Phases sphere cavity are the preserved parts of the Lunar Cylinder lost gearing. Tomographies processed by the authors. *E)* A close-up of the internal area of the Lunar cylinder on Fragment C (photo by the first author, Copyright ©Hellenic Ministry of Culture & Sports/Archaeological Receipts Fund). The white arrows depict the imprint of the lost horseshoe spacer. The black arrow depicts the Lunar phases sphere cavity. The red arrow depicts the half preserved gear-z.

Since the Zodiac month dial is a ring, there is a central circular hole in the Front central area (Allen et al. 2016), with a diameter of about 135mm. This central hole should be filled by the lost structures of gear b1 and the Cover Disc of the gear b1, which is also missing today.

## 4. Clues for the hypothesis of the Planets' rotating spheres on the Antikythera Mechanism. Objections and Contradictions

Many researchers believe that the Antikythera Mechanism had a Planet indication gearing system on the Central Front Face which represented the motion of the 5 planets, but today is lost (Wright 2002; Freeth and Jones 2012; Freeth et al., 2021). This hypothesis is based on four clues which are presented below:

**- Clue I**: On the Back Cover plate of the Mechanism, the engraved *Instruction manual* of the device (Back Cover Inscription) is partially/poorly preserved (Bitsakis and Jones 2016b). The operational parts of the Front plate are presented in the text of BCI Part-1, while the operational parts of the Back plate, are presented on the BCI Part-2.
The names of the 5 known planets during antiquity, or/with their theophoric names are preserved on the text of the BCI Part-1:
[ΕΡΜΟΥ ΣΤΙΛΒΟΝ]**ΤΟΣ** (Hermes Stilbon - Mercury),
**ΑΦΡΟΔΙΤΗ**<Σ> **ΦΩΣΦΟΡΟΥ** (Aphrodite Phosphoros - Venus),
**ΑΡΕΩΣ ΠΥΡΟΕΝΤΟΣ** (Ares Pyroes - Mars),
[ΔΙΟΣ ΦΑ]**ΕΘΟΝΤΟΣ** (Dias/Zeus - Phaethon - Jupiter), and
[ΚΡΟΝΟΥ ΦΑ]**ΙΝΟΝΤΟΣ** (Kronos Phainon - Saturn).

**- Clue II:** The (theophoric) names of the five planets can also be found on the Front Cover inscription, with information for each planet, related to the planet's timed position on the Ecliptic, e.g. **Ο ΔΕ ΦΩ**[ΣΦΟΡΟΣ , **Ο ΔΕ ΦΑΕΘΩΝ ΕΝ . . . . . ΑΠΟΚΑΤΑΣΤΑΣ**[ΕΙΣ, [Ε]**ΣΠΕΡΙΝΟΝ ΣΤΗΡΙΓΜΟΝ** etc.

**- Clue III:** The Fragment D is an unplaced part of the Mechanism. Fragment D consists of a gear with 63 teeth (r1), a circular plate attached to it and an independent oblong/curved plate that are visible in the Tomographies (see Fragment D analysis in Voulgaris et al., 2022). It was suggested that these parts can be related to the gearing of planet Venus, after its engagement with an additional number of hypothetical/non-existed gears (Freeth and Jones 2012; Freeth et al., 2021).

**- Clue IV:** There are preserved mechanical remains fixed on gear b1, which lead to the conclusion that a mechanical structure existed on the b1 gear and extended above it. These lost structures and the lost Cover Disc of gear b1 occupied the space of the large central large hole of the Front plate, **Figure 3**.

Based on these four Clues, the hypothesis of the Planet Indication Gearing on the Antikythera Mechanism appeared through time. It was suggested that the Antikythera Mechanism was a mechanical planetarium device, representing the motions of the five planets and their position on the Ecliptic via their corresponding colored spheres and pointers on the Central Front Face (Wright 2002 and 2013 in bronze model; Freeth et al., 2021-3D computer simulation).

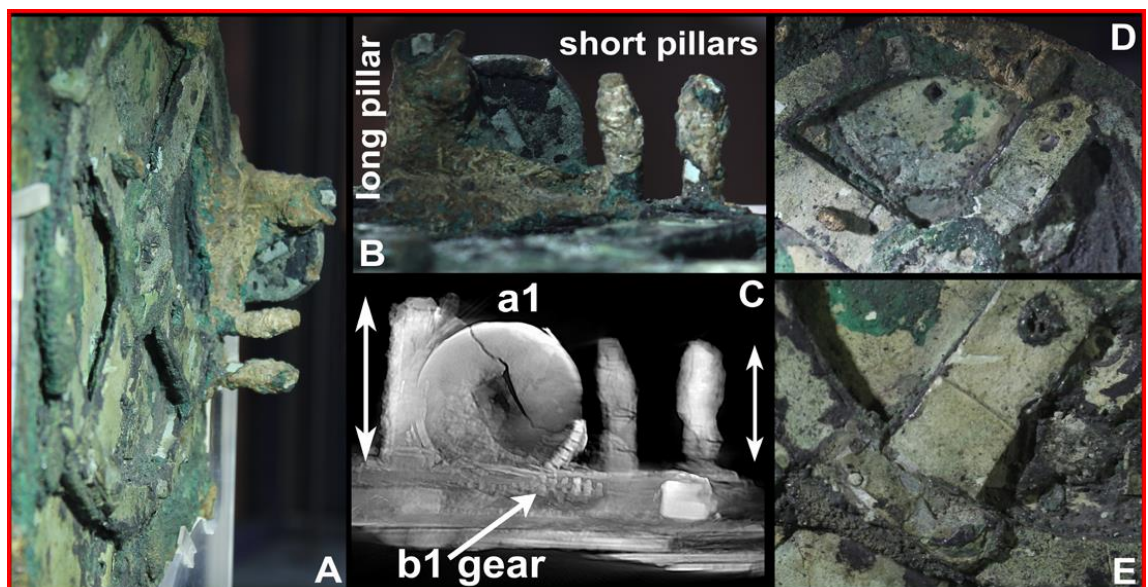

*Figure 3. **A)** Fragment A1 captured in a steep angle (photos by first author, Copyright ©Hellenic Ministry of Culture & Sports/Archaeological Receipts Fund). The four arms of gear b1, the contrate a1 and the one long and two short pillars are visible. **B)** A close-up of contrate a1, the long pillar (on left) and the two short pillars (on right). The long pillar height is larger than the height of the contrate gear and a plate attachment can be done without problematic contact with the gear. **C)** X-Ray combined CTs on same area and scale to panel B. Using the Real3D VolViCon software (from raw volumes), the slices were aligned to the a1 gear surface and were further processed by the authors.*
***D)** The upper part of gear b1. At 45° CCW the small oblong part fixed on the gear's arm is visible.*
***E)** The preserved edge of the thin sheet stabilized with pins on the perimeter of gear b1.*

The authors do not agree with the hypothesis of the Planet Indication Gearing - mechanical Planetarium, for a number of reasons presented and discussed below. Instead of the hypothesis of the planetarium device, the authors will present arguments leading to the conclusion that the Antikythera Mechanism is a Luni-solar time calculator (also taking into account the analysis in Section 2), which had only the Moon and Sun spheres and pointers on the Central Front Face of the Mechanism, as only these two celestial bodies were necessary for Time measurement of the calendar calculation (cycles, years, months) and eclipse predictions during the Hellenistic Era.

The following arguments answer to the above mentioned clues, and contradict the hypothesis of the planet indication gearing on the Antikythera Mechanism:

- **First argument against the planets gearing hypothesis:** A reference of the planet's names is not enough to prove the existence of a Planet Indication gearing. It is possible to reconstruct the text on the Back Cover inscription related to the names of the planets, although the relative text is partially/poorly preserved. In Bitsakis and Jones 2016, p.234, BCI Lines 16-25 (the left/right boundary of the following table corresponds to the Cover plate boundary):

| |
|---|
| 16) **ΠΡΟΕΧΟΝ ΑΥΤΟΥ ΓΝΩΜΟΝΙΟΝ Σ**………………………………………………….. ΠΕΡΙ |
| 17) **ΦΕΡΕΙΩΝ Η ΜΕΝ ΕΧΟΜΕΝΗ ΤΩΙ ΤΗΣ** [………………………….….. ΕΡΜΟΥ ΣΤΙΛΒΟΝ- |
| 18) **ΤΟΣ ΤΟ ΔΕ ΔΙ ΑΥΤΟΥ ΦΕΡΟΜΕΝ**[ΟΝ ……………………………………………… |
| 19) **ΤΗΣ ΑΦΡΟΔΙΤΗ\<Σ\> ΦΩΣΦΟΡΟΥ . . .**[ ………………………………………………….. |
| 20) **ΤΟΥ** [ΦΩ]**ΣΦΟΡΟΥ ΠΕΡΙΦΕΡΕΙΑΝ .**[……………………………………………… |
| 21) **ΓΝΩΜΩ**[.]**ΚΕΙΤΑΙ ΧΡΥΣΟΥΝ ΣΦΑΙΡΙΟΝ . .**[………………………………………… |
| 22) **ΗΛΙ**[ΟΥ] **ΑΚΤΙΝ ΥΠΕΡ ΔΕ ΤΟΝ ΗΛΙΟΝ ΕΣΤΙΝ** [………………………………………… |
| 23) [---ΤΟ]**Υ ΑΡΕΩΣ ΠΥΡΟΕΝΤΟΣ ΤΟ ΔΕ ΔΙΑΠΟΡΕ**[ΥΟΜΕΝΟΝ ……………..……………… |
| 24) [ΔΙΟΣ ΦΑ]**ΕΘΟΝΤΟΣ ΤΟ ΔΕ ΔΙΑΠΟΡΕΥΟΜΕΝΟΝ**[……………………………. Ο ΤΟΥ ΚΡΟ- |
| 25) [ΝΟΥ ΦΑ]**ΙΝΟΝΤΟΣ ΚΥΚΛΟΣ ΤΟ ΔΕ ΣΦΑΙΡΙΟΝ ΦΛ**[……………………………………. |

As the phrase ΚΡΟΝΟΥ ΦΑ]**ΙΝΟΝΤΟΣ ΚΥΚΛΟΣ** (Kronos Phainon circle - Orbit) is preserved, then the word **ΚΥΚΛΟΣ** - Orbit should exist for the planets Mars, and Jupiter.
Therefore the text restoration is:

| |
|---|
| 22) **ΗΛΙ**[ΟΥ] **ΑΚΤΙΝ ΥΠΕΡ ΔΕ ΤΟΝ ΗΛΙΟΝ ΕΣΤΙΝ** [……………………………. ΕΣΤΙΝ <u>Ο ΚΥΚ-</u> |
| 23) <u>ΛΟΣ ΤΟ</u>]**Υ ΑΡΕΩΣ ΠΥΡΟΕΝΤΟΣ ΤΟ ΔΕ ΔΙΑΠΟΡΕ**[ΥΟΜΕΝΟΝ ΣΦΑΙΡΙΟΝ …………<u>Ο ΚΥΚΛΟΣ ΤΟΥ</u> |
| 24) <u>ΔΙΟΣ ΦΑ</u>]**ΕΘΟΝΤΟΣ ΤΟ ΔΕ ΔΙΑΠΟΡΕΥΟΜΕΝΟΝ** [ΣΦΑΙΡΙΟΝ ………… <u>Ο ΤΟΥ ΚΡΟ-</u> |
| 25) <u>ΝΟΥ ΦΑ</u>]**ΙΝΟΝΤΟΣ ΚΥΚΛΟΣ ΤΟ ΔΕ ΣΦΑΙΡΙΟΝ ΦΛ**[……..……………………………….. |

The names of the planets are directly connected to the word **ΚΥΚΛΟΣ** – orbit of planet. At this point, the *ancient Craftsman* presents the names of the planets and the corresponding orbit for each Planet. There is not a clear and distinct presentation for the spheres and the pointers of the planets. The *ancient Craftsman* uses the word **ΚΥΚΛΟΣ**/orbit for each of the superior planets Mars, Jupiter and Saturn and the word **ΠΕΡΙΦΕΡΕΙΑ**/circumference for the orbit of the inferior planets Mercury and Venus (see Bitsakis and Jones 2016, p. 233, lines 17 and 20) and probably for the Sun-Golden sphere.
Moreover, the word ΣΦΑΙΡΙΟΝ (preserved on Line 25 and is implied on Lines 23 and 24) can be well correlated to the ΧΡΥΣΟΥΝ ΣΦΑΙΡΙΟΝ (e.g. ΚΥΚΛΟΣ ΕΣΤΙΝ ΤΟΥ ΑΡΕΩΣ ΠΥΡΟΕΝΤΟΣ. ΤΟ ΔΕ ΔΙΑΠΟΡΕΥΟΜΕΝΟΝ (ΧΡΥΣΟΥΝ) ΣΦΑΙΡΙΟΝ ΕΝ ΣΥΝΟΔΩ BL Z, (there is a circle of Mars.

The rotating Golden sphere is in conjunction (to Mars) in 2 Years + $1/7^y$, see the analysis in https://arxiv.org/pdf/2207.12009 , p. 42-43)

A dominant question arises: *Why did the ancient Craftsman present the orbits of the Planets, instead of their spheres and pointers?*
Each of the planetary orbits was (probably) engraved by a simple circular line on the b1 Cover disc, but it is less important than its sphere and pointer.
The ΣΦΑΙΡΙΟΝ (sphere) and the pointer are operational parts, whereas the ΚΥΚΛΟΣ/orbit does not affect the planet's positional calculation and its existence is not necessary.
A proper text about the planets' sphere/pointer presentation should be ΕΣΤΙΝ ΣΦΑΙΡΙΟΝ ΤΟΥ ΑΡΕΩΣ ΠΥΡΟΕΝΤΟΣ, ΕΠΙ ΚΥΚΛΟΝ ΔΙΑΠΟΡΕΥΟΜΕΝΟΝ, (there is the sphere of Ares Pyroes which travels through its circle) or ΕΣΤΙΝ ΤΟ ΤΟΥ ΚΡΟΝΟΥ ΦΑΙΝΟΝΤΟΣ ΣΦΑΙΡΙΟΝ, ΕΠΙ ΤΟΝ ΑΥΤΟΥ ΚΥΚΛΟΝ ΦΕΡΟΜΕΝΟΝ (there is the sphere of Kronos Phainon etc.), since the *ancient Craftsman* presents the Sun as **ΧΡΥΣΟΥΝ ΣΦΑΙΡΙΟΝ** (Golden sphere) and **ΗΛΙ**[ΟΥ] **ΑΚΤΙΝ** (sun ray-pointer).

From the statistical analysis it results that the *ancient Craftsman* uses (see **Table 2**):
- About seven sentences to describe the Lunar cylinder + the Lunar phases sphere + the lunar pointer + its operation,
- About three sentences (2++) to describe the Golden Sphere-Sun + its pointer + its operation and
- Definitely, only one sentence is used for each of the planets Mars, Jupiter and Saturn (also Mercury and Venus), see **Table 2**.

The rest of the planets should have a similar mechanical design to that of the Sun i.e. spheres in different colors, with pointers (e.g. red for Mars), they should have had a similar text description as the Golden sphere-Sun description and in equal length in the text (as the Sun was one of the planets according to the Hellenistic Astronomy, Toomer 1984, Manitius 1880).

*Table 2.* A part of the preserved Back Cover inscription (Lines 10-26) of the Antikythera Mechanism Instruction Manual (Part-1) (Bitsakis and Jones 2016b, p.232-233). In bold the preserved letters.

| Preserved Inscription | Related to the | Number of sentences |
|---|---|---|
| 10) [- - - - - - - - - -]**ΕΠ ΑΚΡΟΥ Δ**[ ……………………………… <br> 11) [- - - - - - - - - -].**ΩΣΜΕΝΩΝ** .[………………………………… <br> 12) [- - - - - - - - - -]**Ε ΜΕΛΑΝ ΟΤ** .[……………………………… <br> 13) [- - - - - - - - - -]. . . . . . **ΛΩΝΓΕΓ**[……………………………… <br> 14) [- - - - - - - - -].. **Ε** . **ΔΥΠΟΛΑΒΕΙ**[Ν………………………… <br> 15 [. ]**ΟΘΕ**  **ΤΟ ΣΦΑΙΡΙΟΝ ΦΕΡΕ** [……………………………… <br> 16) **ΠΡΟΕΧΟΝ ΑΥΤΟΥ ΓΝΩΜΟΝΙΟΝ Σ**[………………… ΠΕΡΙ | Lunar Cylinder | 7+ |
| 17) **ΦΕΡΕΙΩΝ Η ΜΕΝ ΕΧΟΜΕΝΗ ΤΩΙ ΤΗΣ**[…ΕΡΜΟΥ ΣΤΙΛΒΟΝ- | Lunar Cylinder + Mercury | 1+ |
| 18) **ΤΟΣ ΤΟ ΔΕ ΔΙ ΑΥΤΟΥ ΦΕΡΟΜΕΝ**[ΟΝ……………………………… | Mercury + Venus | |
| 19) **ΤΗΣ ΑΦΡΟΔΙΤΗ<Σ> ΦΩΣΦΟΡΟΥ** . . .……………………… | Venus | 1+ |
| 20) **ΤΟΥ** [ΦΩ]**ΣΦΟΡΟΥ ΠΕΡΙΦΕΡΕΙΑΝ** .[……………………… | Venus + Sun | |
| 21) **ΓΝΩΜΩ**[.]**ΚΕΙΤΑΙ ΧΡΥΣΟΥΝ ΣΦΑΙΡΙΟΝ** . .[……………………… <br> 22) **ΗΛΙ**[ΟΥ] **ΑΚΤΙΝ ΥΠΕΡ ΔΕ ΤΟΝ ΗΛΙΟΝ ΕΣΤΙΝ**…………………… | Sun | 2+ |
| 23) [---ΤΟ]**Υ ΑΡΕΩΣ ΠΥΡΟΕΝΤΟΣ ΤΟ ΔΕΔΙΑΠΟΡΕ**[ΥΟΜΕΝΟΝ… | Mars | 1 |
| 24) [ΔΙΟΣ ΦΑ]**ΕΘΟΝΤΟΣ ΤΟ ΔΕΔΙΑΠΟΡΕΥΟΜΕΝΟΝ**[……(ΚΡΟ) | Jupiter | 1 |
| 25) [ΝΟΥ ΦΑ]**ΙΝΟΝΤΟΣ ΚΥΚΛΟΣ ΤΟ ΔΕ ΣΦΑΙΡΙΟΝ ΦΛ**[………… | Saturn | 1 |
| 26) [- - - - - - -]**ΕΡΑ ΔΕ ΤΟΥ ΚΟΣΜΟΥ ΚΕΙΤΑΙ** . . .[…………………… | | |

The description of the Sun with its Golden sphere and pointer and its operation occupies three lines. But there is not enough space for the description of the rest of the planets with their orbits, spheres, pointers and operation in one sentence per each planet.

**- Second argument against the planets gearing hypothesis:** The preserved text on the Front Cover Inscription (FCI) describes the motion of the planets but there is no reference to a mechanical part related to a planet on the Mechanism e.g. there should be a phrase such as ΓΝΩΜΟΝΙΟΝ ΚΡΟΝΟΥ (Saturn pointer) or ΣΦΑΙΡΙΟΝ ΔΙΟΣ (Sphere of Jupiter) etc.

The total absence of any mechanical reference in the inscription of the Front Cover plate related to the planets' spheres and their pointers can be justified if the Front Cover Inscription was an introductory informative text to the planetary phases, conjunction, morning station, greatest elongation, i.e. the characteristic motions of the planets and their positions on the Ecliptic, as the planets were part of the Cosmos in that era. This text seems somewhat unrelated to the Antikythera Mechanism as there is no a direct connection/relation to parts of the Mechanism.
The specific text is completely independent from the rest and could be a standalone text unrelated to the Antikythera Mechanism.

**- Third argument against the planets gearing hypothesis:** The measurements in Voulgaris et al., 2018b, 2022 (p. 116) and in Roumeliotis 2018, proves that "*the Input of the Antikythera Mechanism is very doubtful from the axis a1, as it creates problems in the mechanical parts, has low torque and makes the handling difficult because of the lack of precision in the pointing*". The central hole of the contrate gear a1 has an oblong shape and an axis with oblong cross section edge, is attached. These two parts could prefix and come from a different machine or construction i.e. a scrap material/useless part, and the *ancient Craftsman* processed them into a gear and axis (as one part).
The argument that "*the Mechanism Input is from gear a1, because it is a common sense assumption from 1974*" cannot be accepted as a scientific argument.

Additionally, when operating the Mechanism from gear a1, the Lunar pointer rotates very fast, making it difficult to aim the Lunar pointer at a desired position with precision (Voulgaris et al., 2022). The design of any mechanical construction that a human will handle, must obey the theory of biomechanics, bioengineering and the motions of the human body. E.g. if the mechanical system of a car's break pedal has a very short travel, it will make it difficult for the driver to control, whereas a very large travel makes proper stopping of the car slow and difficult. Any machine designed to be used by humans must have an easy and efficient operation.

The proper and ideal Input for the Antikythera Mechanism is the Lunar Cylinder/$b_{in}$ axis, as it presents satisfactory torque (Roumeliotis 2018) and also results to a high pointing accuracy of the Lunar pointer to any subdivision of the Zodiac month ring (Voulgaris et al. 2022, p. 116).
The ancient Greek calendars and the time measurement during the Mechanism's era were based on the Lunar synodic month. The Lunar Cylinder, as the Input of the Mechanism, correlates the Mechanism's calculations to the Lunar synodic cycle. Additionally, all of the preserved dials of the Mechanism are mostly based on the lunar synodic cycle.

Today, only three lunar cycles are represented on the Antikythera Mechanism: Sidereal, Synodic and Anomalistic. The fourth, very important, critical and decisive Draconic cycle for the solar eclipses' calculation (well known in the Hellenistic era), is missing. The unplaced Fragment D/gear r1 can satisfactory cooperate with the preserved gears b1-a1-r1 and the hypothetical/lost gear s1. By this engagement, the Draconic cycle is represented and so the four Lunar cycles are present on the Antikythera Mechanism **Figure 4**.

By introducing the Draconic lunar cycle on the Antikythera Mechanism, the lost eclipse events on the Saros spiral can be detected, by the relative positions of the Lunar pointer (to the Golden sphere-Sun or opposite position) and the Draconic pointer (between the ecliptic limits of the Draconic scale), Voulgaris et al., 2023b. For this correlation only one hypothetical small gear of 22 teeth in diameter ≈12mm, is needed (Voulgaris et al., 2022). The scenario that "*the ancient Craftsman presents in his creation the three out of four lunar cycles, he does not present the very important Draconic cycle, and he also presents the five planets*", is inadequate and astronomically incomplete and unjustified.

In Voulgaris et al., 2023b it was shown that the sequence of eclipse events, engraved in the Saros cells, was calculated using the (missing/lost) Draconic gearing on the Antikythera Mechanism.

Freeth et al., 2021 suggests a hypothetical Draconic gearing on the Front face of the Mechanism, without a corresponding dial. In this suggestion the Draconic pointer, is referred as *The Dragon Hand*, gets motion from the b1 gear (via a number of additional gears). The b1 gear rotates in constant angular velocity and therefore this Draconic pointer also rotates with constant angular velocity. But all the lunar cycles have a variable angular velocity, due to the variable velocity of the Moon (Anomalistic cycle). As the lunar motion is variable, the lunar nodes change their position in the sky not steadily, but with variable angular velocity, which is faster when the Moon is near its perigee.

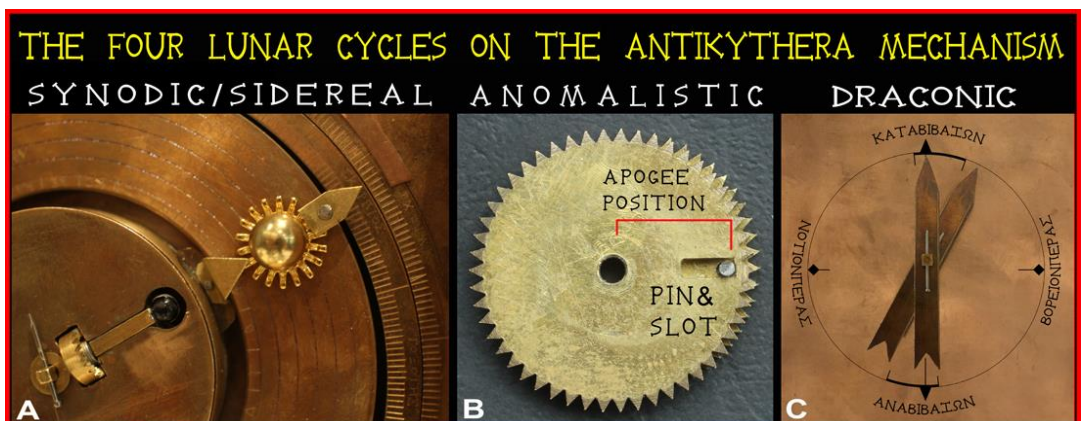

*Figure 4.* The four Lunar cycles present in the Antikythera Mechanism. **A)** The Lunar pointer aims to the Golden sphere-Sun and after New Moon a new Synodic cycle begins. The pointer of the Golden sphere aims to the 13$^{th}$ day of the zodiac month Scorpio and to the 3$^{rd}$ day of the Egyptian month Chilai (in this photo the two rings are not calibrated). The re-aiming of the Lunar pointer to the 13$^{th}$ day of Scorpio corresponds to one Sidereal lunar cycle. **B)** The pin&slot configuration is the "heart" of the Mechanism's timer (Freeth et al., 2006; Wright 2005b; Gourtsoyannis 2010; Voulgaris et al., 2018b and 2023). As Geminus mentions, the Anomalistic cycle begins when the Moon is located at Apogee. The Apogee on the Mechanism occurs when the pin is on its largest distance from the k2 gear axis. **C)** By introducing the Draconic gearing (gears b1-a1-r1/fragment D and s1/hypothetical) all of the known lunar cycles during Antiquity, are presented on the Antikythera Mechanism. The Draconic pointer is attached to gear s1 and it rotates around the Draconic scale in which the two Nodes and their corresponding ecliptic limits are located in opposite positions (Voulgaris et al. 2022 and 2023b).

Therefore, the Draconic pointer should also have a variable velocity. In the suggested model by Freeth et al., 2021, a kind of a *pin&slot* gear design should be added in the suggested gearing, in order to represent this Draconic pointer with variable velocity, as the *ancient Craftsman* designed the preserved *pin&slot* (gears e5/e6/k1/k2) in order to represent the variable motion of the Moon in the sky (Anomalistic cycle).

**- Fourth argument against the planets gearing hypothesis:** The mechanical remains in gear b1 can be related to necessary operation(s) of the Mechanism which are currently missing and without the hypothesis of the Planet Indication Gearing. The analysis and discussion is presented in the next two sections.

## 5. Remains of mechanical parts on the gear b1

The gear b1 is the largest gear of the Mechanism with a diameter of about 129mm. It is not a robust design/one piece bronze disc, as the e3 gear, but it is a combination of a ring with four radial arms distributed by 90°, further increasing the time and the effort for this construction (instead of a simple full metal disc as is the e3 gear) as shown in **Figure 5**. It seems that it was constructed by scrap parts which they left on Craftsman's machine shop. Even today the metal recycling is a common practice by the craftsmen[2].

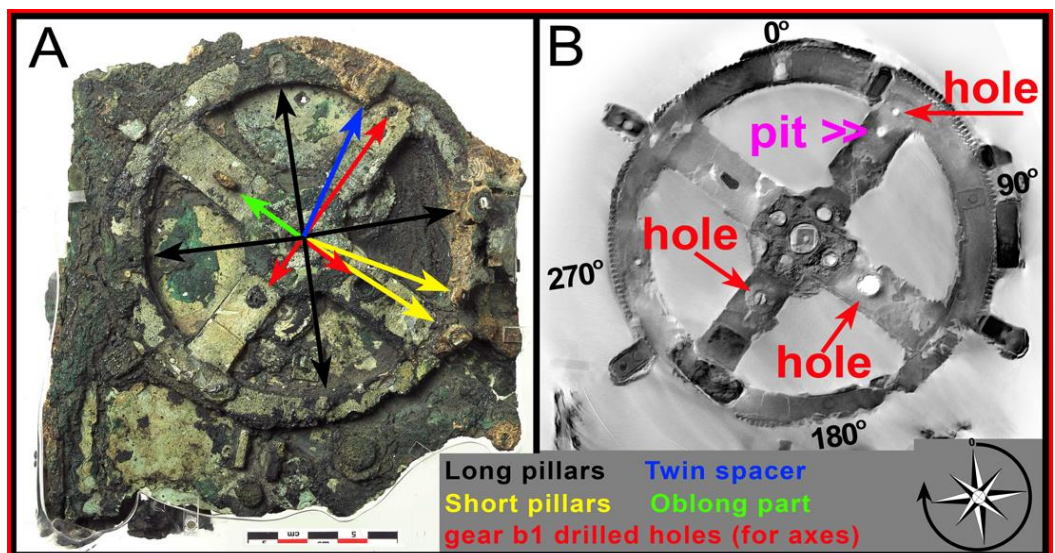

*Figure 5. A)* Visual photograph of Fragment A1 (K. Xenikakis, Copyright ©Hellenic Ministry of Culture & Sports/Archaeological Receipts Fund). Colored arrows in polar coordinates. Black arrows (top black arrow is defined as 0°): the positions of the long pillars. Yellow arrows: the positon of the short pillars. Blue arrow: the one preserved (twin) oblong spacer. Red arrows: the drilled holes, mechanical accepted for axes adaptation. Green arrow: the small oblong part. *B)* Same scale CT-scan of gear b1. The "hole" depicted by the magenta colored symbol ">>" is not a real hole, but it is a hemispherical pit - blind hole. This pit does not fully perforate the gear arm. It seems that it didn't have a mechanical role and just (probably) existed in the raw bronze material before its process, as also the square dug area below the pit. The X-Ray tomographies were processed by the authors, using the Real 3D VolViCon software.

---

[2] M.T.Wright constructed the Middle plate of his model using a bronze recycled scrap plate that it was the name plate from an office door and a pub door kicking plate (Marchant 2009, p. 199). The large, deep engraved letters on the middle plate are visible, see
https://www.youtube.com/watch?v=eC54F4vv_8E at 09:00 and 11:10.

Since bronze was too expensive in that era (also today), it is possible that the *ancient Craftsman* might have attempted to save on this material. E.g. the Middle plate of the Mechanism has much shorter dimensions than the Front/Back plates, as it was not necessary and didn't have mechanical parts (Voulgaris et al., 2019b).

The four arms are stabilized on the ring by the design of the "dove tail" shape, offers a very good stability. Today, the "dove tail" male/female shape design is used for many optomechanical systems' stabilization (telescopes on their mounts, microscope tables, guide sliding tables, in woodworking etc.).

On the b1 gear there are some preserved remains of mechanical parts, shown in **Figures 3, 5 and 6,** which constitute evidence for the existence a number of missing mechanical parts. These remains imply the existence of additional calculation processes performed by the Mechanism that are not preserved. One of these parts, the long pillar, can be directly related to the Cover Disc of gear b1 and the remaining three long pillars should have existed, since their imprints and holes are preserved symmetrically distributed by 90° around the gear's perimeter. On Table 3, the mechanical remains of the gear b1 are presented, described, as also their operation (where this is possible).

**Table 3.** The remains (or their imprints) of the mechanical parts and their position on the gear b1 are presented. A mechanical description and their (probable) operation according to the Personal Constructional Characteristics of the ancient Craftsman (see **Section 7**) are presented.

| Mechanical part on the b1 gear | Description | Starting from b1 gear top arm at −7° | Comments | Operation |
|---|---|---|---|---|
| 1 | **Long pillar (preserved)** | 90° | Height ≈27mm **Figure 3** | $1^{st}$ of 4 pillars for the b1 Cover Disc bearing |
| 2 | Long pillar (non-preserved) | 0° | The imprint of its base is visible **Figure 5** | $2^{nd}$ of 4 pillars for the b1 Cover Disc bearing |
| 3 | Long pillar (non-preserved) | -90° | The imprint of its base is visible **Figure 5** | $3^{rd}$ of 4 pillars for the b1 Cover Disc bearing |
| 4 | Long pillar (non-preserved) | 180° | The imprint of its base is visible **Figure 5** | $4^{th}$ of 4 pillars for the b1 Cover Disc bearing |
| 5 | **Short pillar-1 (preserved) Figure 3** | +120° | Height ≈19mm | For the Thin Sheet Strip-II stabilization, **see Section 8** |
| 6 | **Short pillar-2 (preserved) Figure 3** | +135° | Height ≈19mm | For the Thin Sheet Strip-II stabilization, **see Section 8** |
| 7 | (Twin) **Oblong spacer-1 (preserved) Figure 6** | +40° | Stabilized on the gear perimeter | Spacer for the edge of the Thin Sheet Strip-I bearing See **Section 8** |
| 8 | (Twin) Oblong spacer-2 **(Today not preserved)** Same part as in the previous line | +50° | Spacer's imprint in Rehm and Svoronos photograph, **Figure 6** | Spacer for the edge of the Thin Sheet Strip-I bearing See **Section 8** |

| 9 | **Small oblong part with a hole for a pin adaptation Figures 3A-D** | -45° | **Figures 3, 5, 11, 12, 13** | For the Thin Sheet Strip-II edge immobilization See analysis in **Section 7** |
| --- | --- | --- | --- | --- |
| 10 | **Hole in b1 arm Figures 3D, 5, 6F-G,** | 45° | For a small pillar adaptation **Figures 12, 13A-B** | For the Thin Sheet Strip-I edge immobilization |
| 11 | The dug pothole (pit) in b1 arm | 45° | Origin from scrap material, before the parts process(?) **Figure 6F** | Non-related to mechanical parts |
| 12 | **Edge of a flexible Thin Sheet Strip-I** | -135° | It is stabilized with pins on the b1 gear **Figure 3E** | See its reconstruction in **Section 8**, **Figures 12, 13A-B** |

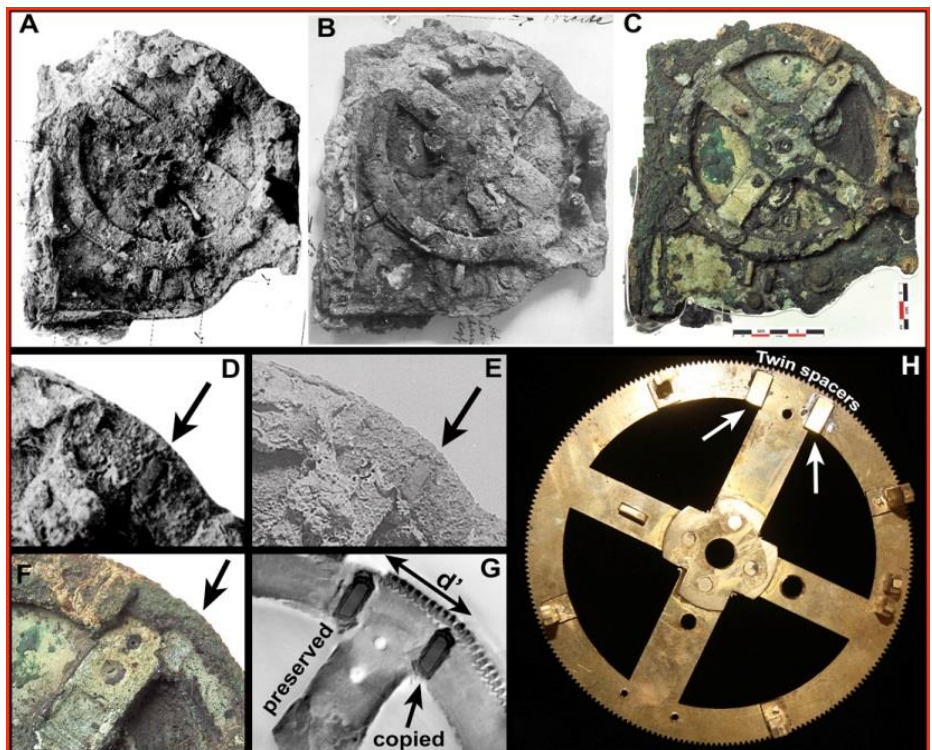

*Figure 6. A)* *Fragment A1 in Svoronos 1903a (photograph period 1902-1903, scan from microfilm in Price's archive, Adler Planetarium, Chicago, https://archive.nyu.edu/handle/2451/59970).* ***B)*** *Fragment A1 in Rehm 1905–1906 (Rehmiana III/9).* ***C)*** *Present day condition of Fragment A1 (photo K. Xenikakis, Copyright ©Hellenic Ministry of Culture & Sports/Archaeological Receipts Fund).* ***D)*** *Close-up of panel A: the second twin spacer is visible in Svoronos 1903a photograph.* ***E)*** *Close–up of panel B: The second twin spacer more clearly visible in Rehm 1905–1906 photograph.* ***F)*** *Close-up of panel C: today, the second twin spacer is missing.* ***G)*** *X-Ray tomography of gear b1, same area as panel F. The second twin spacer was digitally inserted in its corresponding position by copying the preserved twin spacer. The distance d' between the twin spacers defines the width of the lost part (bronze thin sheet, analyzed below) which was in contact to the twin spacers.* ***H)*** *A bronze reconstruction of gear b1. The twin spacers are adapted on their position according to panel G. Also, the four long pillars, the two short pillars and the small oblong part at –45° gear arm position, are constructed and adapted on the gear b1. The part at the b1 center in the shape of a cross seems to have a circular shape, visible in Svoronos 1903a photograph of Fragment A1.*

Even if there were no additional processes related to the b1 gear, these 4 pillars are needed for the b1 Cover Disc attachment (see next section).

At the 45° arm of gear b1, there is a pothole – a dug area. This pothole seems to be unrelated to a mechanical procedure. It could be present in the initial scrap bronze material before its processing to form a gear arm. The attachment of a moving part or its base in this pothole by the use of glue is not in accordance to the *Personal Constructional Characteristics* of the *ancient Craftsman* (see further below) and the use of the glue for moving parts' axes/shafts is mechanically risky and doubtful.

## 6. The *b1 Cover Disc* of the Antikythera Mechanism's Front face

In terms of engineering and manufacturing, the Antikythera Mechanism is created by a professional craftsman of that era. It should therefore be at the limit of further improvement, comprising every necessary procedure for its operation and present a constructional uniformity: parts used for the same tasks should have similar design and assembly style. Additionally, every part must exhibit maximum design efficiency and there should not exist any non-essential or unjustified parts.

The user of the Mechanism should not be concerned about the internal parts. E.g. all of the internal mechanical parts of a car, especially the moving parts, are not visible to the driver: e.g. above the gear box there is the gear lever. All of the gear levers are covered by a leather or plastic flexible conical cover for mechanical protection and aesthetical reasons. This design prevents any insertion unwanted elements inside the moving parts of the gear box. This way the gear lever and box appears as a complete construction.

A cover should exist above the gear b1, for protection (to prevent stray parts from entering inside) and also for aesthetic reasons (from now on we call this cover as the *b1 Cover Disc*). Additionally, the surface of the *b1 Cover Disc* offers a satisfactory space for measuring scale(s) or other information.
The *b1 Cover Disc* should follow the internal round shape of the Zodiac month ring. The *Cover Disc* can be stabilized on the four long pillars of the b1 gear. The height of the preserved long pillar is larger than the diameter of the crown gear a1, which is engaged to the b1 gear (see **Figure 3**) and therefore the *b1 Cover Disc* can be based on the four long pillars without a mechanical malfunction.
The *b1 Cover Disc* rotates with the same angular velocity as the b1 gear i.e. 1 full turn/solar tropical year and it is the only external part of the Mechanism with a period of one Tropical year. Therefore the Golden Sphere-Sun should be directly fixed on the Cover disc via a small pillar. This design is in agreement to the reconstructed phrase by the authors of the preserved description for the Golden sphere-Sun:
…ΕΙΣ **ΓΝΩΜΩΝ ΚΕΙΤΑΙ**. **ΧΡΥΣΟΥΝ ΣΦΑΙΡΙΟΝ** ΕΠΙ ΓΝΩΜΩΝΑ ΕΣΤΙΝ "… lies a gnomon. A Golden sphere is attached on the pillar" (in bold the preserved text, Bitsakis and Jones 2016b, p. 233, lines 20-21; see the mechanical scheme in Voulgaris et al., 2019b; see also https://arxiv.org/pdf/2207.12009, p. 13, 31).
The (lost) base of the Lunar Cylinder would be in contact to the *b1 Cover Disc* in order to have a good stability during its rotation.

About 25% of the 365 subdivisions (days) of the solar tropical year cycle are preserved on the Zodiac month ring. The solar pointer **ΗΛΙ**ΟΥ **ΑΚΤΙΝ**/Sun ray (Bitsakis and Jones 2016) travels through the subdivisions. Based on the *Personal Constructional Characteristic No 4* of the *ancient Craftsman* (*each pointer is accompanied by its corresponding dial*, see next section), the 29½ subdivisions for the Lunar synodic cycle could be engraved in circular distribution on the b1 Cover Disc, in circular dimension, right after the Lunar Cylinder.[3] This scale shows the age of the Moon, as the Lunar pointer travels along it. In the BCI a phrase is preserved in Line 16: **ΠΡΟΕΧΟΝ ΑΥΤΟΥ ΓΝΩΜΟΝΙΟΝ Σ**[ΕΛΗΝΗΣ [a little (Lunar) pointer projected from it… see Bitsakis and Jones 2016b, p.232].

Based on the preserved inscription ΚΡΟΝΟΥ ΦΑ]**ΙΝΟΝΤΟΣ ΚΥΚΛΟΣ**, Saturn orbit-circle, six concentric cycles, each for the rest of the planets, should follow right after the 29½ subdivisions (days), representing the orbits of Mercury, Venus, Sun, Mars, Jupiter and Saturn (see the analysis and the tables in **Contradiction argument I**).

## 7. The *Personal Constructional Characteristics* of the ancient

Any creator designs and constructs his creation according to his/her experience, aesthetics, and manual dexterity. The combination of these personal constructional skills creates the constructional style of the creator, which is specific and personal. There is always a characteristic "imprint" of the creator in every construction. In many cases, it is possible to identify the creator by observing the construction style, similar to identifying a composer by the style of his musical composition (or the era of that composition).

As the Antikythera Mechanism was designed and constructed by an experienced Craftsman, its creation presents a number of distinctive **Personal Constructional Characteristics** (**PCC**). By the study the AMRP X-Ray tomographies of the Fragments, we detected these Constructional Characteristics, which are the "*constructional imprint*" of *the ancient Craftsman*.

### 7.1 *PCC 1*: The stabilization of the moving parts

For the stabilization of the moving parts on their axes i.e. gears and pointers, which are parts that they have a torque/load, the *ancient Craftsman* used stabilizing pins, attached perpendicularly to the parts' axes as shown in **Figure 7**.

---

[3] Many old cathedral clocks present the Lunar Age dial, see http://www.patrimoine-horloge.fr/as-padoue.html ; http://www.patrimoine-horloge.fr/as-brescia.html ; http://www.patrimoine-horloge.fr/as-clusone.html.

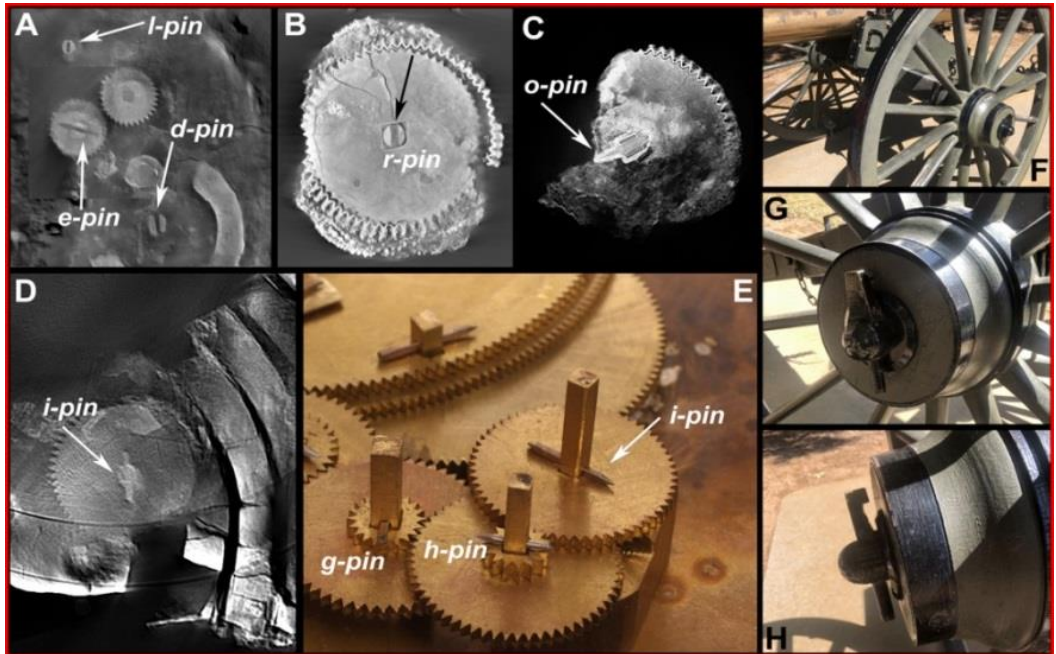

*Figure 7. A, B, C, and D) selected X-Ray tomographies depicting the stabilizing pins perpendicular to the axes/shafts d, e, l (Fragment A), r (Fragment D), o (Fragment B) and I (Fragment A2). The tomographies were processed by the authors (based on the Raw volumes). This is the Personal Constructional Characteristic No 1 of the ancient Craftsman. E) Close-up of the authors' functional model, showing the way of gears g, h, i stabilization according to the Personal Constructional Characteristic No 1 of the ancient Craftsman. F, G, H) A similar way of stabilizing of a cannon wheel, as a rotated disc on its axis using a perpendicular pin (cannon in front of the Texas State Capitol). Photo by the first author.*

The Craftsman could have used glue for bronze (alloy of lead/tin, see Voulgaris et al., 2018c) in order to stabilize the gears on their corresponding shafts. But this would be risky, since at any time it could come off, halting the Mechanism. He could also apply pressure with a small (wooden) hammer to fix the gears on their shafts.

Eventually the ancient Craftsman used the safest, for that era, way to stabilize the gears on their corresponding shafts by the use of a perpendicular pin, in contact to the upper surface of the gear. This also offers a practical way to remove and reposition the parts of the Mechanism.

**7.2 *PCC 2*:  The Ω-plates and the thin sheet strips**
The *Craftsman* of the Mechanism, in order to stabilize an axis (and a gear), when using a thin sheet/bar, made an oblong hole on the edge of the sheet, then made an oblong part of equal shape and dimensions as the oblong hole and finally fixed that part to its base **Figure 8 and 9**.

For the stabilization of the thin sheet, he adapted the thin sheet via its oblong hole on the oblong fixed part. Then he secured the thin sheet by using a pin perpendicular to the oblong part. In this design, the pin(s) are in a direction parallel to the base.

This style of stabilization is clearly detected on the *Ω-plate* for the restraint of the d-shaft (see Voulgaris et al., 2018c, Fig. 8b). The *Ω-plate* is stabilized on the Middle plate, via two oblong parts fixed on the Middle Plate and two pins, penetrating the oblong parts perpendicularly **Figure 8**.

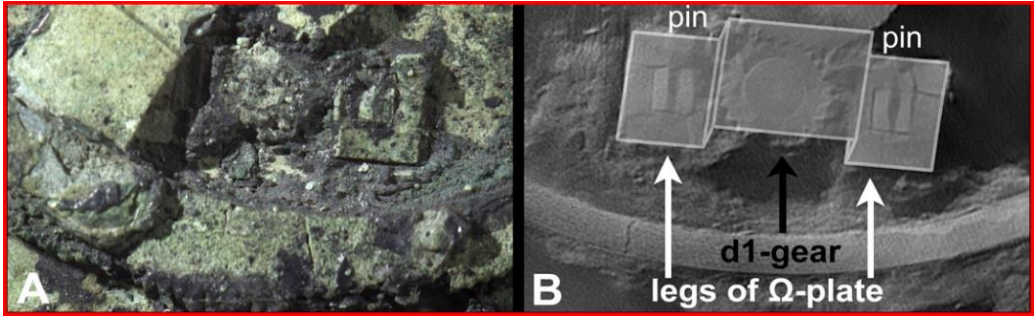

*Figure 8. A) The Ω-plate of shaft-d is partially preserved on Fragment A1. The right Ω-"leg", the small oblong part and its stabilizing pin are visible. Photo by the first author, Copyright ©Hellenic Ministry of Culture & Sports/Archaeological Receipts Fund. B) X-Ray tomography of the same area to panel A, processed by the authors using Real3D VolViCon software (using the AMRP Raw volume). The shape design of the lost Ω-plate of shaft d was sketched (see also Voulgaris et al., 2019b).*

Also evident, although partially preserved, is the *Ω-("butterfly shape") plate* for the restraint of the gears e5/e6 and k1/k2 on their axes/shafts. This *Ω-plate* is stabilized on the e3 gear via two oblong parts (today only one is preserved) and two pins parallel to the gear's surface, as shown in **Figure 9** (see also Fig. 12 and 13 in Wright 2005b).

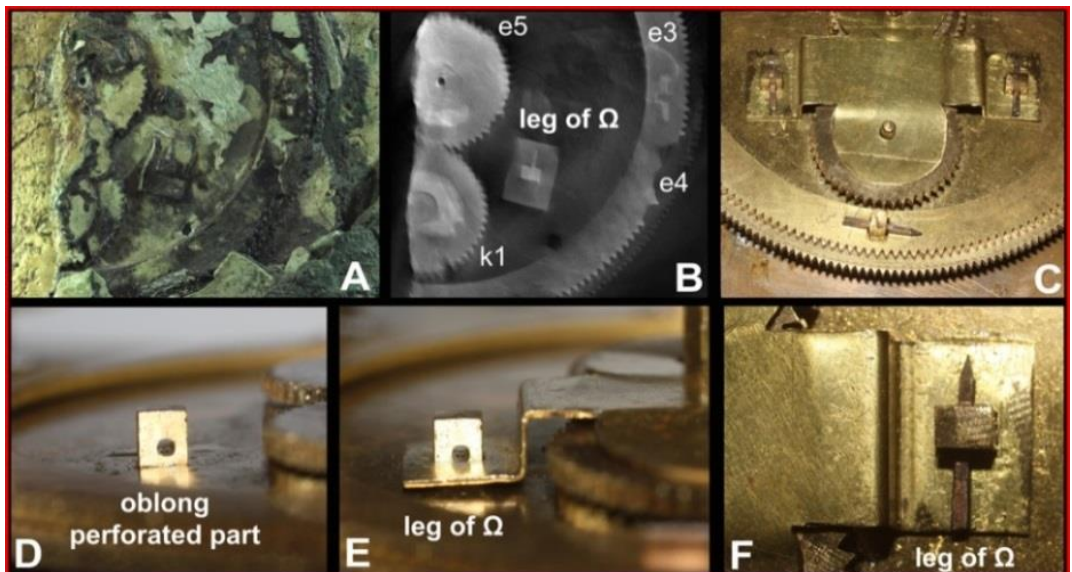

*Figure 9. A) The half-preserved gears e3, e4, e5, e6, k1 and k2, on Fragment A2. One "leg" of the lost Ω-plate is also preserved with its stabilizing pin (parts of Ω-plate are preserved on the gears e6 and k2). The Ω-plates are the Personal Constructional Characteristic No 3 of the ancient Craftsman (photo by the first author, Copyright ©Hellenic Ministry of Culture & Sports/Archaeological Receipts Fund). B) X-Ray tomography of the same area to panel A, processed by the authors using Real3D VolViCon software (using the AMRP raw volume). The CT-slice is aligned to the e3 level. C) A close-up of the reconstructed Ω-plate for gears e6/e5 and k2/k1 and the gears e3 and e4. The Ω-plate was reconstructed using a thin bronze sheet.. D) The oblong part is fixed on the gear e3. All of the preserved small oblong parts of the Mechanism have a hole for the stabilization of the Ω-plate "leg" via a pin. E) The "leg" of the Ω-plate is adapted on the small oblong part. F) Afterwards, the Ω-"leg" is secured with a pin inserted to the hole of the small oblong part.*

Also, the stabilization of the gear e4 on e3 gear is done with the use of the four small oblong perforated parts (today only one is well preserved) fixed on the gear e3, four oblong holes on gear e4 and four stabilizing pins, perpendicularly adapted on the oblong part and parallel to the gear's surface, **Figure 10**.

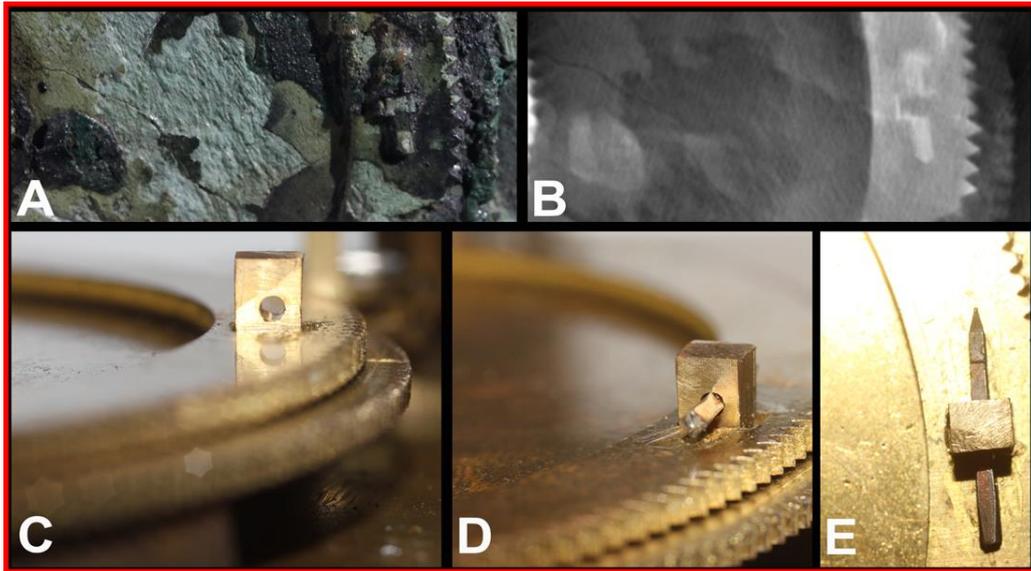

*Figure 10. A)* The stabilizing pin for securing gear e4 on gear e3 (one out of four is preserved on Fragment A2. Photograph by the first author, Copyright ©Hellenic Ministry of Culture & Sports/Archaeological Receipts Fund. *B)* X-Ray tomography of the same area of panel A. The slice was aligned to the surface of e3 gear using Real3D VolViCon software (from AMRP raw volume). *C)* The oblong small part which is fixed on gear e3 and its hole, according to the X-Ray tomographies (see also Wright 2005b). *D* and *E)* The pin inserted in the hole of the small oblong part to secure gear e4 on gear e3, as it is in contact to gear e4 surface. Parts' reconstruction and photographs by the first author.

The preserved small oblong part located on the b1 gear' arm at –45° has a similar design to all the preserved small oblong parts, **Figure 11**. Therefore, the operation of the small oblong part of the b1 gear, should be similar to the operation of the rest oblong parts (see Figure 9), i.e. a thin bronze sheet, should be inserted in this oblong part and a pin should be secure this strip.

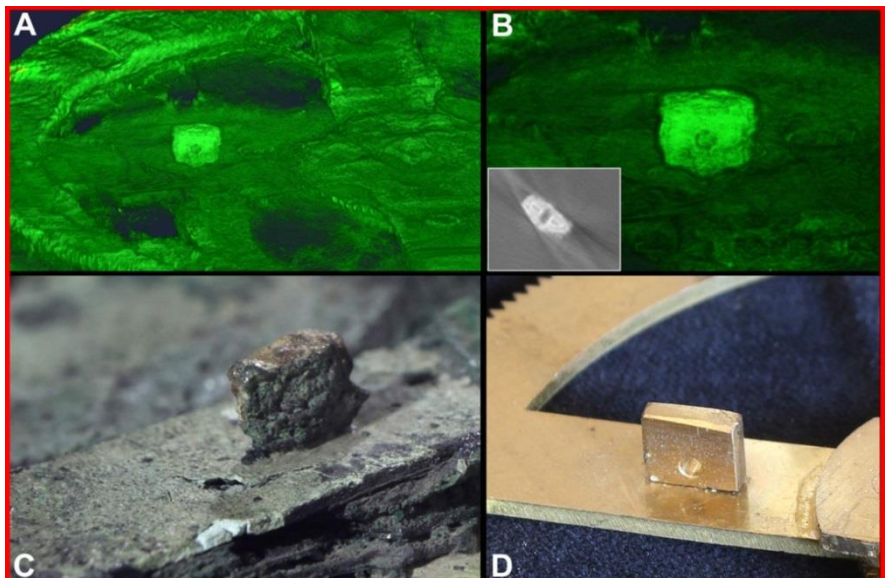

*Figure 11. A)* The small oblong part at –45° arm is fixed on the b1 gear. This part has a hole for a pin insertion (see also Freeth and Jones 2012). *B)* Close-up of panel A. The small oblong part and its hole are visible Insert: view of the small oblong part from top. The 3D reconstructions after the Raw Volumes process by the authors using Real3D VolViCon software. *C)* The oblong part as preserved today (Photograph by the first author, Copyright ©Hellenic Ministry of Culture & Sports/Archaeological Receipts Fund). *D)* Bronze reconstruction of the oblong part adapted on a bronze b1 gear (bronze parts reconstructions and photo by the first author).

### 7.3 *PCC 3*: Using Spacers for the gears' support

The *ancient Craftsman* used spacers to better stabilize the gears by increasing their bearing surface. In this way, a gear rotates on a constant plane: About all of the Mechanism's gears have been constructed by a simple bronze plate 2mm–2.5mm thick. Since all the gears are thin they cannot remain totally perpendicular to their axes/shafts, and therefore deviate from perpendicularity to their corresponding axes/shafts (Voulgaris et al., 2018b, 2018c and 2022 Fig. 12A; Wright 2002).

The ancient Craftsman inserted a number of spacers which acted as bases for the gears, improving their functionality. All the spacers are fixed somewhere (mostly to the middle plate), are non-rotating parts, do not suffer from the torque/load and they do not have tension or strain. Therefore, their stabilization can be done with pins or glue, suited for bronze without affecting the mechanical motion of the system (Voulgaris et al., 2018c).

### 7.4 *PCC 4*: The pointers rotated on their calibrated dial

The ancient Craftsman paired every pointer with a corresponding dial: All of the Antikythera Mechanism pointers rotated on their corresponding calibrated dials. A pointer without its calibrated dial does not present any measurement. All of the Mechanism's dials are subdivided with a constant unit period (subdivisions per turn): equal subdivisions for the circular dials (one year/subdivision on the Games dial, one Saros/subdivision in Exeligmos dial, and gradually increasing dimension (but in constant period), for the cells of the two spiral dials Metonic and Saros (1 synodic cycle/cell).

Hence, the ratio *Pointer's Angle/Subdivision = constant number.*

These four basic constructional characteristics signify the *Personal Constructional Characteristics* of the *ancient Craftsman* of the Antikythera Mechanism.

We applied the four **Personal Constructional Characteristics** of the *ancient* in all of our functional models and in our suggested reconstructions (described below), in order to keep a high degree of the originality and the quality in our research Our constructional protocol forbids the stabilization of a gear or a moving part using modern stabilizing parts (e.g. screws, nuts etc.) or glue. The study of the **Constructional Characteristics** of the *ancient Craftsman* gave us the ability to see and to experience the way the *ancient Craftsman* was thinking for the parts' assembly and what problems he faced during the parts' construction, their assembly and use of his creation. We have also detected the constructional and measurement limits of the Mechanism, which differ from the theoretical study or the 3D simulations (Voulgaris et al., 2023b).

### 8. Reconstructing lost mechanical parts of the b1 gear by applying the *Constructional Characteristics of the ancient Craftsman*

The remains on the b1 gear are poorly preserved, but by applying the **four *Constructional Characteristics of the ancient Craftsman***, the reconstruction of the lost parts of the b1 gear can be achieved, as some of the mechanical remains of gear b1, have similar or identical design characteristics to other parts, which are better preserved and whose operation we already know.

**1)** The 1+3 long pillars can be related to the attachment/stabilization of the b1 Cover Disc, which is the outer central front face of the Mechanism, as sown in **Figure 6 and 14**.

**2)** One preserved spacer is placed at 40° and the imprint of the second similar spacer at 50° (well visible in Svoronos 1903a and in Rehm 1905–1906 photographs). We call these two similar spacers, as twin spacers (see **Figure 6**). In the opposite position (-135°), there exists one edge of a thin flexible bronze sheet. The second (lost) edge of this sheet should be in contact to the twin spacers, and therefore the thin sheet extends to the whole diameter of the b1 gear, as shown in **Figure 12**. The stabilization of the *Thin Sheet Strip-I* can be done with a perpendicular pin, according to the *Constructional Characteristic 2.*

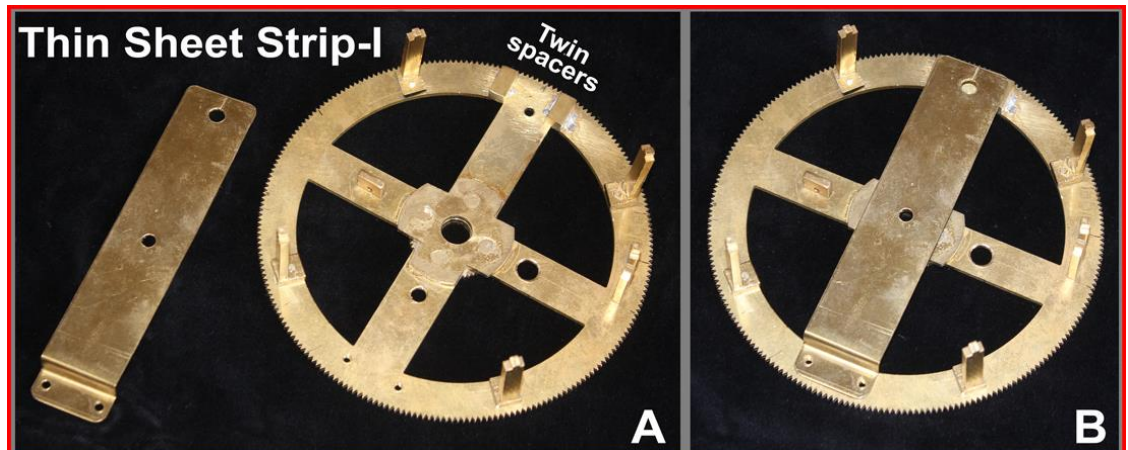

*Figure 12. A) Bronze reconstruction of the (flexible) Thin Sheet Strip-I and the b1 gear. The width of the Thin Sheet Strip-I was designed according to the preserved part located at the edge of 225° gear arm (see Figure 3E). B) Adapting the thin sheet strip I on the b1 gear. The lower edge of the bronze strip is retained by pins on the b1 gear according to the preserved part of the original prototype. The upper edge of the sheet is in contact to the Twin spacers according to PCC3. The upper edge is stabilized with the use of a stabilizing pin on a (necessary) hypothetical pillar. There is a hole preserved on the b1 gear (at the 45° arm and between the twin spacers), in which a pillar can be attached (see next figure).*

**3)** According to the *Constructional Characteristic No 2*, the small oblong part with a hole located at the –45° position on the b1 arm, should be the stabilizer for the edge of a second flexible thin bronze sheet. The edge of the thin sheet is stabilized with the use of a perpendicular pin in the hole of the small oblong part. On the opposite diameter (at 120° and 135°), there are two short pillars. The second edge of the thin sheet is stabilized on the two short pillars via two perpendicular pins, as shown in **Figure 13**. This stabilization method is in perfect accordance to the constructional characteristics of the thin sheets of the *Ω-plates.*

According to our suggested reconstruction of the b1 gear lost parts, they are two different thin sheet strips in crossed direction, located at different heights/levels above b1 gear **Figure 14**. The height of the twin spacers (≈6mm) defines the height of the *Thin Sheet Strip-I* above the b1 gear surface and the height of the two short pillars (≈16mm) defines the height of the *Thin Sheet Strip-II*.

.

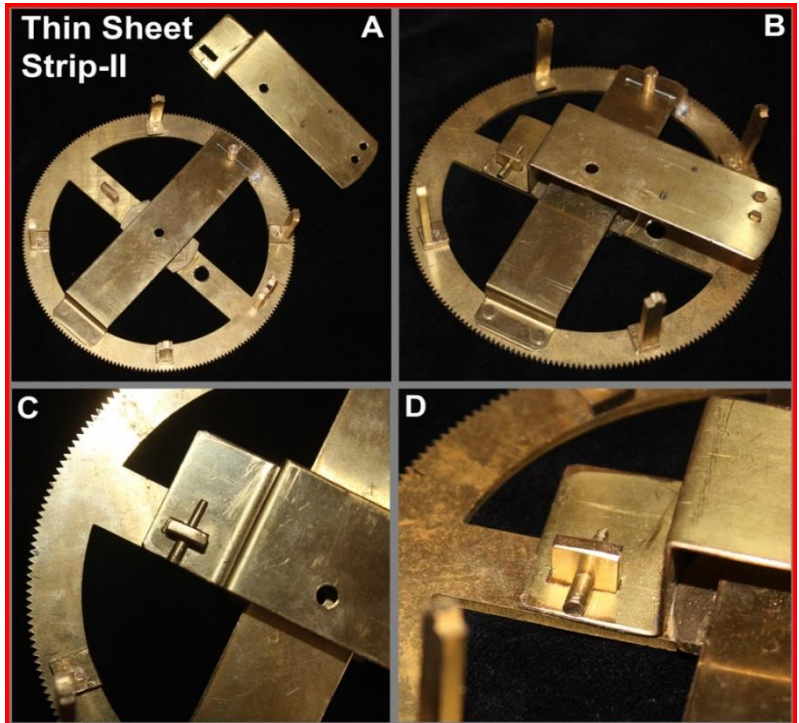

*Figure 13. A) The design of the Thin Sheet Strip-II. B) Stabilizing the Thin Sheet Strip-II on the gear b1 according to the PCC3. C and D) The way of the thin sheet strip stabilization via the perpendicular pin. The stabilization via the pin is the same as in the preserved Ω-plates (see Figures 8 and 9).*

The Cover Disk should be the place where pointers and corresponding dials associated with additional necessary procedures were located (see next section).

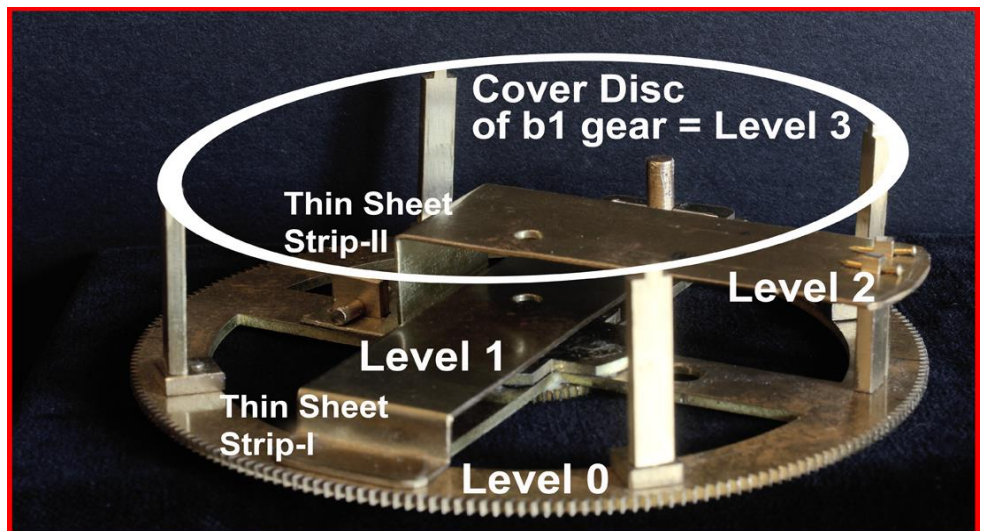

*Figure 14. In our suggested reconstruction there are four levels in the gear b1. Level 0: the b1 gear's upper surface. Level 1: the Thin Sheet Strip-I. Level 2: the Thin Sheet Strip-II. Level 3: the position of the Cover Disc for gear b1.*

## 9. Measuring procedures that must have been present in the AM but are not preserved.

Two necessary measurement procedures which are not preserved today on the Mechanism emerged during the extended use of our functional models of the Antikythera Mechanism. The absence of these two procedures creates difficulties and doubts while using the Antikythera Mechanism: many times, the operation had to be stopped, to find the time point corresponding to the pointers. The problem was even worse when only one user was handling the Mechanism. In order to avoid losing count during the Mechanism's operation, we used some small sticky notes, attached to the Mechanism's front plate.

It is evident that this is not a correct and proper way for operating the Antikythera Mechanism and makes its' use problematic, irregular and difficult. This practice negates the device's independent and self-contained operation.

Both of the necessary - but missing from its current state – procedures are directly related to the Front Central Dial of the Mechanism and they can be introduced on the Mechanism via gearings, pointers and scales, taking into account the existing holes and the mechanical remains of the b1 gear and the b1 Cover Disc.

**9.1 First necessary (not preserved) measurement process:**

The Front central face of the Mechanism (along with the two parapegma plates) is dedicated to the solar tropical year: the Ecliptic sky - Zodiac month ring (Fragment C), in which the Golden sphere - Sun and its pointer travel around the ring of the Zodiac constellations. One full turn of the solar pointer (ΗΛΙΟΥ ΑΚΤΙΝ/Sun-ray, see Bitsakis and Jones 2016b), corresponds to one solar tropical year. The 19 solar tropical years of one Metonic cycle, are equal to 19 full rotations of the Golden sphere-Sun.

On the Back plate of the Mechanism, the Metonic spiral consists of 235 cells/synodic months of the Metonic cycle. Each of the 19 unequal Metonic years is a repeated set of 13 or 12 months as Geminus mentions (Manitius 1880; van der Waerden 1984b; Fotheringham 1924; Theodosiou and Danezis 1995). Therefore, 19 unequal Metonic years correspond to 19 equal solar tropical years.

The position of the Metonic pointer aiming at a specific cell of the Metonic spiral, defines the current Metonic year: in the first month/cell (ΦΟΙΝΙΚΑΙΟΣ/ Phoinikaios month) of each Metonic year, the *ancient Craftsman* engraved successively the symbols **LA, LB, LΓ … LIΘ** (Year 1, 2, 3,… 19), which correspond to the numbered Metonic years (Freeth et al., 2008; Anastasiou et al., 2016). The Metonic years are in disharmony to the solar tropical years and each Metonic year precedes the solar tropical by about 3–28 days (i.e. max ≈95% of the dodecatemorion).

When the user operates the Mechanism by turning the Input-Lunar Cylinder (Voulgaris et al., 2022 and Roumeliotis 2018), he has no information of how many turns the Golden sphere Sun has completed (1 tropical year equals 13.368 full turns/Sidereal cycles, of the Lunar Cylinder). Therefore, while operating the Mechanism, the user either should keep in mind the number of the Golden sphere full rotations (but losing count on many occasions), keep notes (as the authors did), or he has to halt the operation of the Mechanism, turn it back to check the Metonic pointer position, and see in which Metonic month/cell of the spiral does the pointer aim. Then he must find the first month *Phoinikaios* by following the order of months in reverse direction and read the engraved symbol L(*number*). Afterwards, the user looks again on the front and continues the operation of the Mechanism.

It is imperative for the Antikythera Mechanism's user to know at any time, very fast and without any interruption, the number of turns completed by the Golden sphere-Sun is (i.e. how many tropical years passed after the initial starting date, see Voulgaris et al., 2023a). This procedure must be located at the Front central face of the Mechanism.

A measurement of the tropical solar years via a pointer and calibrated dial, counting the number of turns of the Golden sphere-Sun around the Ecliptic, is missing from the Antikythera Mechanism Front plate (as the Front plate is dedicated to the Solar Tropical year). The *19 solar tropical year gearing* can be fitted on the b1 gear, the *Thin Sheet Strip-II* and *the Cover Disc*.

The gearing must satisfy the condition for "*a full rotation of the b1 gear, the pointer rotates 1/19 of a turn on its dial*".

Three pairs of gears must satisfy the ratios of the equation:

$(1/2.5)*(1/2)*(5/19) = (1/19)$ or $(1/2)*(1/3)*(6/19) = 1/19$ or $(1/2.5)*(1/2.8)*(7/19) = (1/19)$ or **$(1/2.5)*(1/2.4)*(6/19) = (1/19)$** (Eq. 1) *(adopted in this work)*, see **Table 4**.

**Table 4:** The teeth number of the six suggested gears (**I-VI**) for the *19 Solar Tropical year gearing*.

| The 19 Solar Tropical year gearing on the b1 gear | | | | |
|---|---|---|---|---|
| Numbered gear (I-VI) and their teeth number | **I**: 24 teeth | **III**: 20 | **V**: 18 | |
| | **II**: 60 teeth | **IV**: 48 | **VI**: 57 | **= 1/19** |
| Ratio | 1/2.5 | 1/2.4 | 6/19 | |

The *"19 Solar Tropical year Scale"* is divided in 19 equal numbered sectors (Α, Β, Γ, …., ΙΘ) and it is engraved on the b1 Cover Disc, as shown in **Figure 15**.

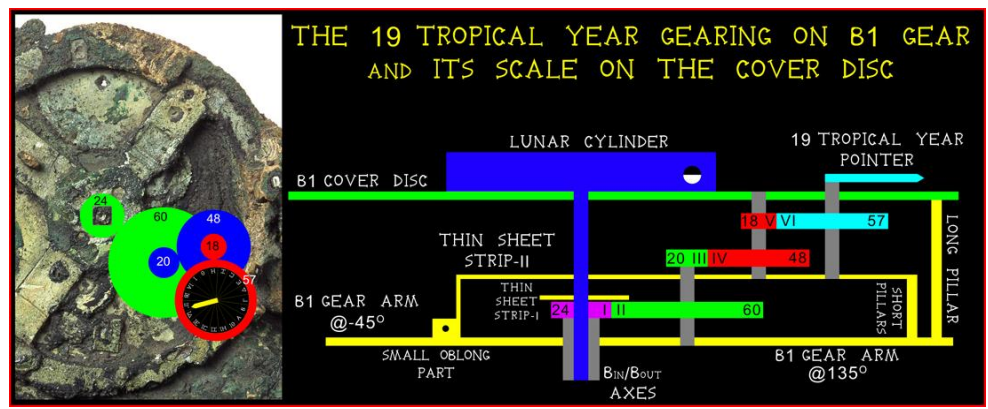

*Figure 15. The gearing scheme which presents the 19 Tropical year measurement procedure. The teeth number of the six gears, their dimension and positioning was calculated according to the preserved position of the holes and the boundary of b1 gear, the Lunar Cylinder dimension and the b1 gear Cover Disc. The off-axis placement of the two preserved short pillars is well justified by the existence of the off-axis placement of the gears IV-V. The Lunar Cylinder and its axis $b_{in}$ are sketched in blue color.[4]*

**9.2 Second necessary (not preserved) measurement process:**

The solar tropical year lasts 365.25 days. The *ancient Craftsman* divided the Zodiac ring in 365.25 subdivisions or rounded in 365.0 (Voulgaris et al., 2018a). The Egyptian calendar of 365.00 days is represented on the Mechanism by the Egyptian calendar ring divided in 365 subdivisions. Both rings are free to rotate (Voulgaris et al. 2018a). Additionally, beneath the Egyptian calendar ring he constructed a third ring with 365 holes of ≈0.8mm diameter, in circular distribution. This process is an arduous work, which demands special care and concentration (Voulgaris et al., 2019a). It is thus evident that the *ancient Craftsman* wanted his sophisticated Mechanism to be very accurate and so he paid a special attention to the Egyptian calendar measuring procedure.

---

[4] At the edge of preserved $b_{in}$ axis ($b_{in}$ axis in blue color) a conical hole is preserved (Voulgaris et al., 2018b). We suppose that this conical hole was created by a careless researcher at the beginning of the last century, in order to place the edge of a compass for some measurements. If this hole really existed in the ancient prototype, then the Lunar Cylinder and its mechanical connection to the Fragment A must be radically reconsidered!

For every four rotations of the Golden sphere-Sun, the user must rotate the Egyptian ring by one subdivision CCW. In this way the Egyptian calendar precedes by one day every 4 tropical years.

During the use of our functional models, very often we forgot to rotate the Egyptian calendar ring CCW after four rotations of the Golden sphere, or after losing count of the Golden sphere rotations we weren't sure if it was the time to rotate the Egyptian calendar ring.

This problem was more pronounced if the Mechanism was not used for some days - it was nearly impossible to remember if the ring has been already turned or not. Before restarting the measuring procedure, we tried to calculate the Golden sphere rotations, and measure the number of subdivisions on the Egyptian ring from the fiducial line mark (Voulgaris et al., 2023a), i.e. from the initial calibration date. Then we rotated (or not) the ring to its proper position but having a doubt for the correct position.

We also tried to correct any counting error either by using small yellow sticky note pads or operating the Mechanism in pairs: one was *the Operator* and the other *the Observer/Counter* for the Golden sphere rotation and the Egyptian ring rotation!
It is evident from the above that this is not a correct, regular and continuous operational procedure for the Antikythera Mechanism.

Therefore, a pointer with a dial alerting the user when it is necessary to rotate the Egyptian calendar ring by one subdivision CCW" is mandatory. The **"Egyptian calendar reminder scale"** can be engraved on the b1 Cover Disc, divided in four quadrants corresponding to the four Tropical years **Figure 16**.

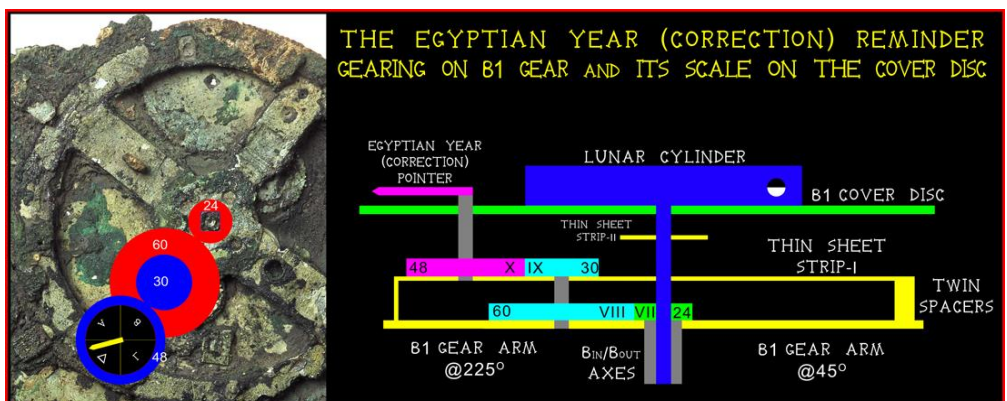

*Figure 16. The gearing scheme which presents the "Egyptian calendar correction reminder" procedure adapted on the gear b1 of Fragment (K. Xenikakis). The teeth number of the four gears, their dimension and positioning was calculated according to the preserved position of the holes and the boundary of b1 gear, the Lunar Cylinder dimension and the b1 gear Cover Disc.*

The gearing should satisfy the ratios of the equation (1/2)*(1/2) = (1/4) or **(1/2.5)*(1/1.6) = (1/4)** (adopted in this work for reasons of gearing uniformity with the previous necessary procedure, see *Eq. 1*), so that the pointer rotates by 1 turn every 4 years see **Table 5**. When the pointer of this dial aims to the letter **Δ** (4), the user should rotate the Egyptian calendar ring by one day/subdivision.

**Table 5:** The teeth number of the four suggested gears (**VII-X**) for the *Egyptian year correction reminder gearing*.

| The Egyptian year correction reminder gearing on the b1 gear ||||
|---|---|---|---|
| Numbered gear (***VII-X***) and their teeth number | **VII**: 24 teeth | **IX**: 30 | |
| | **VIII**: 60 teeth | **X**: 48 | **= 1/4** |
| Ratio | 1/2.5 | 1/1.6 | |

# 10. Discussion: Authors' reflections and doubts about the hypothesis of the planet indication gearing

The hypothetical/suggested planet gearing indication on the Antikythera Mechanism demands the use of an extremely large number of hypothetical/non-preserved mechanical parts (about 60+): gears, axes, shafts, plates, bars, spacers, small and large diameter discs, bearing bases, tubes, large diameter rings, pointers, spheres, side supports etc.
The adaptation of these hypothetical parts definitely increases the static friction and the kinetic friction, the weight, and the inertia to a large degree. These hypothetical parts, apart from being too heavy to hold (about 1.5-2kg), are difficult to assemble, have increased inertia, they are of doubtful functionality, and introduce mechanical problems and non-seamless motion. Of course, if modern design parts, like screws, and other special materials are used for their reconstruction, then they can be functional (see below).
Some of the suggested/hypothetical parts are not in accordance to the **Personal Constructional Characteristics** of the *ancient Craftsman*.
Moreover, there is no clear reference about planets, their spheres and pointers in the *Instruction Manual* of the Mechanism.

The use of the modern mechanical assembly parts such as screws and grub screws, nuts for the parts' stabilization (instead of perpendicular pins that as are used by the *ancient Craftsman*), polished iron/steel axes (instead of bronze), special alloys gear material (that are much harder than the typical bronze), thicker gears (the max thickness of the gears is ≈2.5 mm) or different design gears (the gear design is a simple disc plate, not gear with hub) and the modern involute tooth shape (instead of triangular tooth shape as the original gears), improves the Mechanism's models functionality. However, all these deviations constitute an unacceptable and incorrect practice for the Antikythera Mechanism's research, because they change dramatically the *mechanical status* of the system, leading to different mechanical behavior of the device than the original prototype, and leading to different and wrong conclusions.
One of the most characteristic examples of the *mechanical status* change after adapting different parts is the almost fatal airplane disaster after the use of wrong-sized bolts (2.5mm shorter than the proper size!) in the pilot's window (https://www.newscientist.com/article/mg13418180-300-wrong-bolts-sent-pilot-into-the-blue/).
Using 3D simulations can also be a helpful research tool for the Mechanism study, but it cannot substitute for a physical model. The 3D representations present an ideal and sterilized condition and mechanical behavior, without friction, inertia, constructional mismatches on the gears, periodic errors, non-perfect transmission of motion etc. (Voulgaris et al., 2023b).

Many times we use simple and straightforward ideas to design something based on thought, but as the "*The devil is in the details*" real world designs often display very different behavior than the desired one.

An accepted - research quality - reconstruction model of the Antikythera Mechanism, must be strictly constructed according to the following principles:

1. Be based on the CT scans,
2. Without any change of the parts' ordered position,
3. Without making holes where nothing exists on the ancient original prototype,
4. Keeping the *Personal Constructional Characteristics of the ancient Craftsman*,
5. Without any use of modern mechanical parts,
6. Having gears with triangular teeth shape constructed from a simple bronze plate, according to the CT-scans,
7. Having gear thickness around 2mm,
8. Using a simple bronze alloy material (and not special or ferrous alloys), and
9. Using the minimum hypothetical parts adaptation.

In this way, the Antikythera Mechanism will reveal its "*secrets*".

## 11. Conclusions

In this work we analyzed and presented a probable reconstruction of the mechanical remains located on the b1 gear and the b1 Cover Disc of the Antikythera Mechanism. The mechanical remains can be related to the two missing but necessary procedures of the Mechanism, in order to be considered as a complete and independent time measuring device. The two procedures we suggest, the *19 Solar Tropical year Dial* and the *Egyptian calendar (correction) Reminder Dial* are closely related to the solar tropical year, which is presented by the *ancient Craftsman* on the Front Central area of his creation. According to our experiments and following the use of our models and the problems in the measurements that arose, these two suggested procedures are necessary for the orderly/non-stop operation of the Antikythera Mechanism, as shown in **Figure 17**.

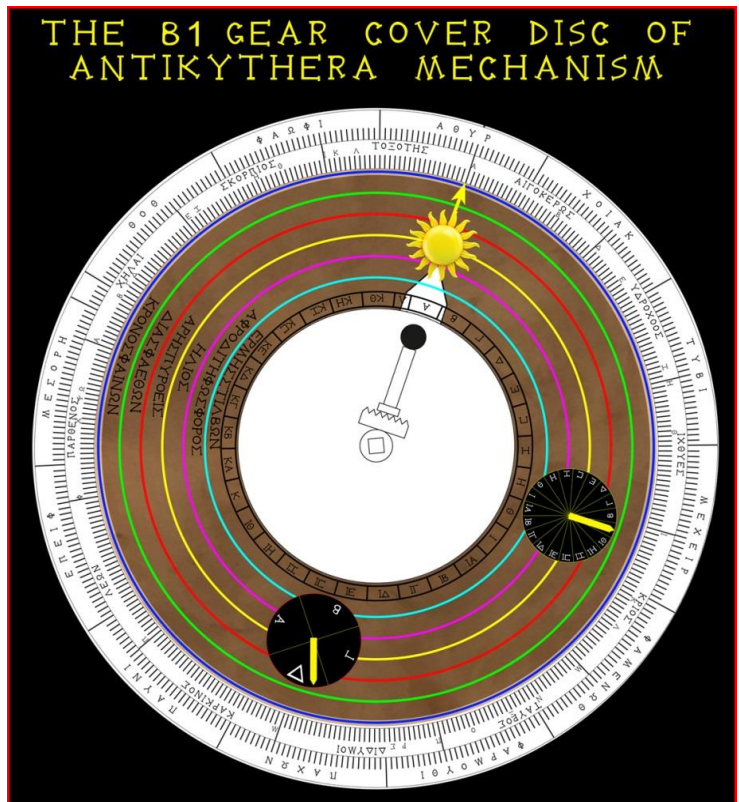

*Figure 17. The reconstructed b1 Cover Disc and its digital placement on Fragment A. The Lunar day dial right after the Lunar cylinder, the Egyptian calendar (correction) Reminder Dial, the 19 Tropical year Dial and the Golden sphere-Sun with its pointer (ΗΛΙΟΥ ΑΚΤΙΝ), are presented. The homocentric ΚΥΚΛΟΙ/orbits of the five planets are also sketched here in different colors. By applying the two necessary/lost procedures on the Antikythera Mechanism, the user can at any time find out in which tropical year does the Golden sphere aim and when it is time to adjust the Egyptian calendar ring. The two outer rings of the Egyptian calendar and the Zodiac month, the Golden sphere pointer, the Lunar pointer and the Lunar phases sphere, are aligned and calibrated according to the initial starting date of the Mechanism's pointers 22/23 December 178 BC (1$^{st}$ day of Capricorn – 17/18 Hathyr, Voulgaris et al., 2023a)*

The existence of the mechanical remains on the b1 gear is not a proof for the hypothetical planet indication gearing and is not the only available suggestion.

The experience we had during the fragments and their CT-scans study, the design, the construction and the extended use of our Antikythera Mechanism functional models has revealed the characteristics of this remarkable machine, the possible problems faced by the *ancient Craftsman* during the parts' construction and assembly and the mechanical limits of this geared system. These limits have an impact on the calculations made by the gears (Voulgaris et al., 2023b).

As many parts and information are missing from this device, the safest way to learn more about the additional procedures of the Mechanism would be to find additional lost fragments.

Someone may think that "*the Antikythera Mechanism seems incomplete or that is lacks completeness without the planets and by only presenting the Sun and the Moon on its central front face*". The answer is presented in **Figure 18**: many clocks in south-western Europe (also in the volvelles of the Medieval era, see Apian 1524; Drennan 2012) present on their face the Sun and the Moon (without planets) and they bear a striking resemblance to the Mechanism design, which has resulted from the research presented in this paper.

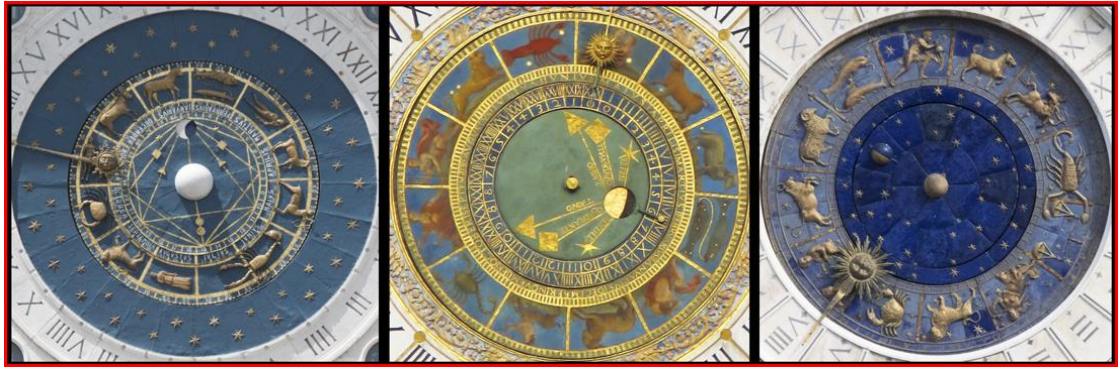

*Figure 18.* *Astronomical clocks in Italy, Padova, Brescia and Venice. The three clocks present similarity in their design to the Antikythera Mechanism: The Earth is located at the center, they have a zodiac ring, the Sun is presented with a pointer, the Moon appears with a lunar phases sphere (or disc) and a Lunar Age scale. Credit and copyright: Anthony Ayiomamitis. Used with permission.*

After a very long time spent in constructing, assembling, handling, testing, studying and interacting with our Antikythera Mechanism functional reconstruction models, our experience and the new observations of the present work, we can conclude that the Antikythera Mechanism was a mechanical geared calculator for calendrical/astronomical/time measuring procedures, based on the relative timed positions of the Moon and the Sun, constructed around 180 BC and must have been constructed for use by an authority for the management of Time in that era.

## Acknowledgements

*We are very grateful to prof. X. Moussas (National and Kapodistrian University of Athens University), who provided us with the X-ray Raw Volume data of the Mechanism fragments and Dr. F. Ullah for his support in the use of the REAL3D VolViCon software. Thanks are due to the National Archaeological Museum of Athens, Greece, for permitting us to photograph the Fragments of Antikythera Mechanism. Many thanks to the photographer A. Ayiomamitis give us the permission to use his photographs of the old clocks from Padova, Brescia and Venice.*